\documentclass[twocolumn]{openjournal} 

\usepackage{orcidlink} 
\usepackage[T1]{fontenc}
\usepackage{ae,aecompl}
\usepackage[T1]{fontenc}
\usepackage{ae,aecompl}
\usepackage{hyperref}
\usepackage{url}

\usepackage{longtable}
\usepackage{booktabs}
\usepackage{tabu}
\usepackage{graphicx}
\usepackage{multirow}
\usepackage{ulem}
\usepackage{color}
\usepackage{amsmath}

\def \s{~\rm{s}}
\def \km{~\rm{km}}

\def \erg{~\rm{erg}}

\def \pc{~\rm{pc}}
\def \kpc{~\rm{kpc}}

\def \keV{~\rm{keV}}

\usepackage{xcolor}
\definecolor{redak}{rgb}{0.9,0.15,0.05}

\shorttitle{Point symmetric structure in CCSNR N132D}
\shortauthors{Soker}

\graphicspath{{./}{figures/}}

\begin{document}

\title{Attributing the point symmetric structure of core-collapse supernova remnant N132D to the jittering jets explosion mechanism}

\author{Noam Soker\,\orcidlink{0000-0003-0375-8987}} 
\affiliation{Department of Physics, Technion Israel Institute of Technology, Haifa, 3200003, Israel}

\date{\today}

\begin{abstract}
 I identified a point-symmetric morphology in the core-collapse supernova (CCSN) remnant (CCSNR) N132D, composed of two symmetry axes: the short symmetry axis extending from the northwest ear and through the center of the iron-rich emission on the other side, and the second along the long dimension of N132D and coincides with the extension of the central oxygen-rich gas to the northeast. Namely, the point-symmetry of the outer zones of CCSNR N132D correlates with that of the oxygen-rich gas near the center. The surrounding gas cannot shape the inner oxygen-rich material, implying that the point-symmetric morphology is a property of the explosion mechanism, as predicted by the jittering jets explosion mechanism (JJEM). The oxygen-rich material is known to be in a torus. According to the JJEM, an energetic pair of opposite jets, more or less perpendicular to the plane of the torus, has shaped the torus; this pair is along the short symmetry axis. Another energetic pair, perpendicular to the first one, shaped the elongated, large-scale structure of CCSNR N132D. I discuss how the JJEM accounts for two perpendicular pairs of jets and the unequal jets in each pair. CCSNR N132D is the fifteenth CCSNR with an identified point-symmetric morphology attributed to the JJEM. Because the neutrino-driven mechanism cannot explain such morphologies, this study further strengthens the claim that the JJEM is the primary explosion mechanism of CCSNe. 
\end{abstract}


\section{Introduction} 
\label{sec:intro}

Studies in the last two years have been discussing two competing core-collapse supernova (CCSN) explosion mechanisms, the delayed neutrino explosion mechanism (e.g., \citealt{Andresenetal2024, BoccioliFragione2024, Burrowsetal2024kick, JankaKresse2024, vanBaaletal2024, WangBurrows2024, Bambaetal2025CasA, Bocciolietal2025, EggenbergerAndersenetal2025, Huangetal2025, Imashevaetal2025, Laplaceetal2025, Maltsevetal2025, Maunderetal2025, Morietal2025, Mulleretal2025, Nakamuraetal2025, SykesMuller2025, Janka2025, ParadisoCoughlin2025, Vinketal2025, WangBurrows2025}) and the jittering jets explosion mechanism (JJEM), which is the frame of the present study.

The JJEM asserts that the primary explosion mechanism of the majority, and likely all, CCSNe involves pairs of jets (e.g., \citealt{Soker2010, PapishSoker2011}). The magnetorotational explosion mechanism is much older, but significantly differs from the JJEM. In the magnetorotational explosion mechanism, one pair of jets along a fixed axis explodes the star. To maintain a fixed angular momentum axis along which the accretion disk launches a pair of jets, the progenitor of the CCSN must be rapidly rotating (e.g., \citealt{Shibagakietal2024, ZhaMullerPowell2024, Shibataetal2025} and references to much older papers therein). The requirement for a rapidly rotating pre-collapse core makes this mechanism rare. Therefore, the magnetorotational explosion mechanism attributes most CCSNe to the neutrino-driven mechanism, and I classify it as part of the neutrino-driven mechanism. 

According to the JJEM, $N_{\rm 2j} \simeq 5-30$ pairs of jittering jets that intermittent accretion disks (or belts) around the newly born neutron star (NS) launch on a timescales of $\tau_{\rm ex} \approx 0.1-10 \s$ explode all CCSNe (for a list of the parameters of the JJEM see \citealt{Soker2025Learning}); time is measured from shock bounce, i.e., the formation time of the shock at the newly born NS. In rare cases when the pre-explosion core is rapidly rotating, the jittering of the pairs of jets is at very small angles around the pre-explosion angular momentum axis. Practically, it is a fixed-axis explosion, some of which end with a black hole remnant. The variations in the axes of the pairs of jittering jets might be fully or partially stochastic.
The two jets in a pair might substantially differ in their opening angle and power (e.g.,  \citealt{Bearetal2025Puppis}), and might be at an angle smaller than $180^\circ$, namely, not exactly opposite (e.g., \citealt{Shishkinetal2025S147}).

Angular momentum fluctuations in the convection zones of the collapsing core that instabilities above the NS (for instabilities turbulence in the gain region, see, e.g., \citealt{Abdikamalovetal2016, KazeroniAbdikamalov2020, Buelletetal2023}) amplify cause the stochastic variations in the angular momentum axis directions of the accreted gas that launches the jets along these axes  (e.g., \citealt{GilkisSoker2014, GilkisSoker2016, ShishkinSoker2021, ShishkinSoker2023, WangShishkinSoker2024}). Neutrino heating plays a role by adding energy to the jets \citep{Soker2022nu}. However, most of the energy originates from the jittering jets, which serve as the primary driver of the explosion. 

In a recent review \citep{Soker2024UnivReview}, I summarized the state of debate between these two alternative explosion mechanisms as of 2024. The main challenge of the JJEM is to demonstrate with three-dimensional numerical simulations the formation of intermittent accretion disks during the explosion process. Such simulations should be with very high resolution, include magnetic fields, and the correct pre-collapse core convection. The next step of launching the jets by these intermittent accretion disks seems to be beyond the present capabilities of numerical simulations \cite{Soker2025Learning}. 
The neutrino-driven mechanism encounters problems and difficulties that the community often overlooks. I list four issues.

(1) There are qualitative and quantitative disagreements between simulations of different research groups (e.g., \citealt{Janka2025}), like on which stellar models explode and which do not. 

(2) Some modelings of CCSNe deduce explosion energies of $E_{\rm exp} \gtrsim 2 \times 10^{51} \erg$ (e.g., \citealt{Moriyaetal2025}), more than the neutrino-driven mechanism can supply. Many magnetar models of superluminous CCSNe require explosion energies of $E_{\rm exp} \gtrsim 3 \times 10^{51} \erg$ that imply explosion by jets (e.g., \citealt{SokerGilkis2017, Kumar2025}), a conclusion overlooked by most magnetar modelers. 

(3) The neutrino mechanism predicts that some massive stars fail to explode, in contradiction with studies that suggest that there is only a small or no population of failed CCSNe (e.g., \citealt{ByrneFraser2022, StrotjohannOfekGalYam2024, BeasoretalLuminosity2025, Healyetal2025}). \citet{Bocciolietal2025} suggested that the neutrino-driven mechanism can overcome this difficulty. 

(4) The most severe and challenging problem the neutrino-driven mechanism encounters is explaining the point-symmetric morphologies of CCSN remnants (CCSNRs; Section \ref{sec:PointSymmetry}). This problem likely rules out the neutrino-driven mechanism as the primary mechanism of the explosion.  On the other hand, the JJEM predicts that many CCSNRs exhibit point-symmetric morphologies, as recent three-dimensional hydrodynamical simulations demonstrate \citep{Braudoetal2025}. For this reason, identifying point-symmetric CCSNRs is of high significance in determining the explosion mechanism. In Section \ref{sec:N132D}, I claim that SNR N132D is another CCSNR with a point-symmetric morphology that supports the JJEM.  

Because many researchers are unaware of these difficulties or purposely ignore them in papers and review talks, the community overrates the neutrino-driven mechanism, as I discuss in the Summary of this study (Section \ref{sec:Summary}).  

\section{Identifying point-symmetry by eyes}
\label{sec:PointSymmetry}

The pairs of opposite jets that explode the star in the JJEM might shape pairs of structural features in the descendant CCSNR. I emphasize that in most CCSNe these jets are not relativistic (\citet{Guettaetal2020} suggest that most CCSNe have no relativistic jets, contrary to gamma ray bursts that do have, e.g., \citealt{Izzoetal2019, AbdikamalovBeniamini2025}). 
Only a small number of pairs leave imprints on the descendant CCSNR; the other explode the core, lose their symmetry, and leave no clear geometrical imprint. Because the symmetry axes of the pairs of jets change their directions from one jet-launching episode to the next, the CCSNR's pairs of structural features do not share the same axis, resulting in a point-symmetric morphology.  The opposite structural features might include dense clumps, dense elongated structures termed filaments, bubbles, which are faint structures closed and encircled by a brighter rim, lobes, which are bubbles with partial rims, and ears, which are protrusions from the main CCSNR shell with decreasing cross-sections away from the center.

Studies identified point-symmetric morphologies in 14 CCSNRs and discussed these in the frame of the JJEM. Some of these CCSNRs possess clear point-symmetrical morphological features, like 
N63A \citep{Soker2024CounterJet}, the
Vela CCSNR (\citealt{Soker2023SNRclass, SokerShishkin2025Vela}), 
Cassiopeia A \citep{BearSoker2025}, and the 
Crab Nebula \citep{ShishkinSoker2025Crab}, 
and others with less secure point-symmetric morphologies, like  
CTB~1 \citep{BearSoker2023RNAAS}. 
   Other CCSNRs and the studies that attributed their morphologies to the JJEM are 
SNR 0540-69.3 \citep{Soker2022SNR0540},
the Cygnus Loop \citep{ShishkinKayeSoker2024},
SN 1987A \citep{Soker2024NA1987A, Soker2024Keyhole}, 
G321.3–3.9 \citep{Soker2024CF, ShishkinSoker2025G321},
G107.7-5.1 \citep{Soker2024CF},
W44 \citep{Soker2024W44}, 
Puppis A \citep{Bearetal2025Puppis},  
SNR G0.9+0.1 \citep{Soker2025G0901}, and 
S147 \citealt{Shishkinetal2025S147}.  
In some CCSNRs, the dust also presents point-symmetric morphologies \citep{Soker2025Dust}. 

Several processes are likely to disrupt the point-symmetric morphology, such as instabilities and the NS natal kick velocity that occur during the explosion, as well as certain post-explosion processes. The latter include interaction with a circumstellar material (CSM) lost by the CCSN progenitor (e.g., \citealt{Chiotellisetal2021, ChiotellisZapartasMeyer2024, Velazquezetal2023, Meyeretal2022, MeyerDetal2024}), interaction with the interstellar medium (ISM; e.g.,  \citealt{Wuetal2019, YanLuetal2020, LuYanetal2021, MeyerMelianietal2024}), pulsar wind nebula if it exists, and heating processes such as reverse shock and radioactive decay. 
In many CCSNRs, the processes that smear the point-symmetric structures induced by the explosion make the identification of point-symmetric morphologies challenging or even impossible.

In this study of the SNR N132D and the earlier studies of point-symmetric CCSNRs, I have been utilizing my over thirty years of experience in classifying planetary nebulae by inspecting their jet-shaped morphologies. The visual-inspection classification of morphologies is a powerful common practice in classifying planetary nebulae (e.g., \citealt{Balick1987, Chuetal1987, Sahaietal2007, Sahaietal2011}), and AGN jets (e.g., \citealt{Hortonetal2025}), and has led to significant breakthroughs, particularly in establishing the major role of binary interaction in the shaping of planetary nebulae, and the role that jets play in many of these interacting binaries.  
 
Although the visual-inspection method might seem completely subjective because it is not quantitative, it is not; it has a large objective component. Humans possess a remarkable ability to critically examine symmetry and identify departures from it by inspection alone, as facial symmetry is a key indicator of mate quality (e.g., \citealt{Rhodes2006, Pinheiroetal2023} and reference therein;). In particular, our ability to identify \textit{fluctuating asymmetries}, which are non-directional (random) deviations from perfect symmetry in bilaterally paired traits, is crucial as fluctuating asymmetries tend to reflect health problems (e.g., \citealt{Rhodes2006}). The majority of authors (and referees) made one of the most important decisions of their lives, if not the most important, i.e., choosing a mate, primarily through visual inspection, particularly by looking for symmetry and ruling out large fluctuating asymmetries. Recognition of symmetry is relatively automatic and consistent across cultures (e.g., \citealt{Rhodes2006}).    

\section{The point symmetric structure of SNR N132D} 
\label{sec:N132D}

Numerous papers study the structure and morphology of SNR N132D (e.g., \citealt{Lasker1978, Lasker1980, Hughes1987, Blairetal1994, DickelMilne1995AJ, Morseetal1995, Morseetal1996, Blairetal2000, Beharetal2001, Tappeetal2006, Tappeetal2012, XiaoAndChen2008, Bambaetal2018, Lawetal2020, Shardaetal2020, Suzukietal2020, Banovetzetal2023, Rhoetal2023, Xrism2024N132D, Fosteretal2025, LongX2025, Okadaetal2025}). None of these studies examined the point-symmetric morphology and structure in the context of the JJEM, which is the goal of this study. Some studies have mentioned a possible jet as an explanation for the runaway knot in the southeast (e.g., \citealt{VogtDopita2011}), which I will refer to later.  

\cite{Morseetal1995} first identified a point-symmetric structure of the fast O-rich ejecta of SNR N132D (Section \ref{subsec:Morse1995}); they did not term it a point-symmetric structure and did not connect it to jet shaping. Before turning to their results, I visually inspect some X-ray images to identify a global point-symmetric morphology (Section \ref{subsec:Visually}). In Section \ref{subsec:3D}, I discuss the three-dimensional (3D) structure from \cite{VogtDopita2011} and \cite{Lawetal2020} in the frame of the point symmetric morphology and the JJEM.   

\subsection{The X-ray point-symmetric morphology} 
\label{subsec:Visually}

\begin{figure*}[]
	\begin{center}
\includegraphics[trim=0.0cm 9.8cm 0.0cm 0.1cm ,clip, scale=0.90]{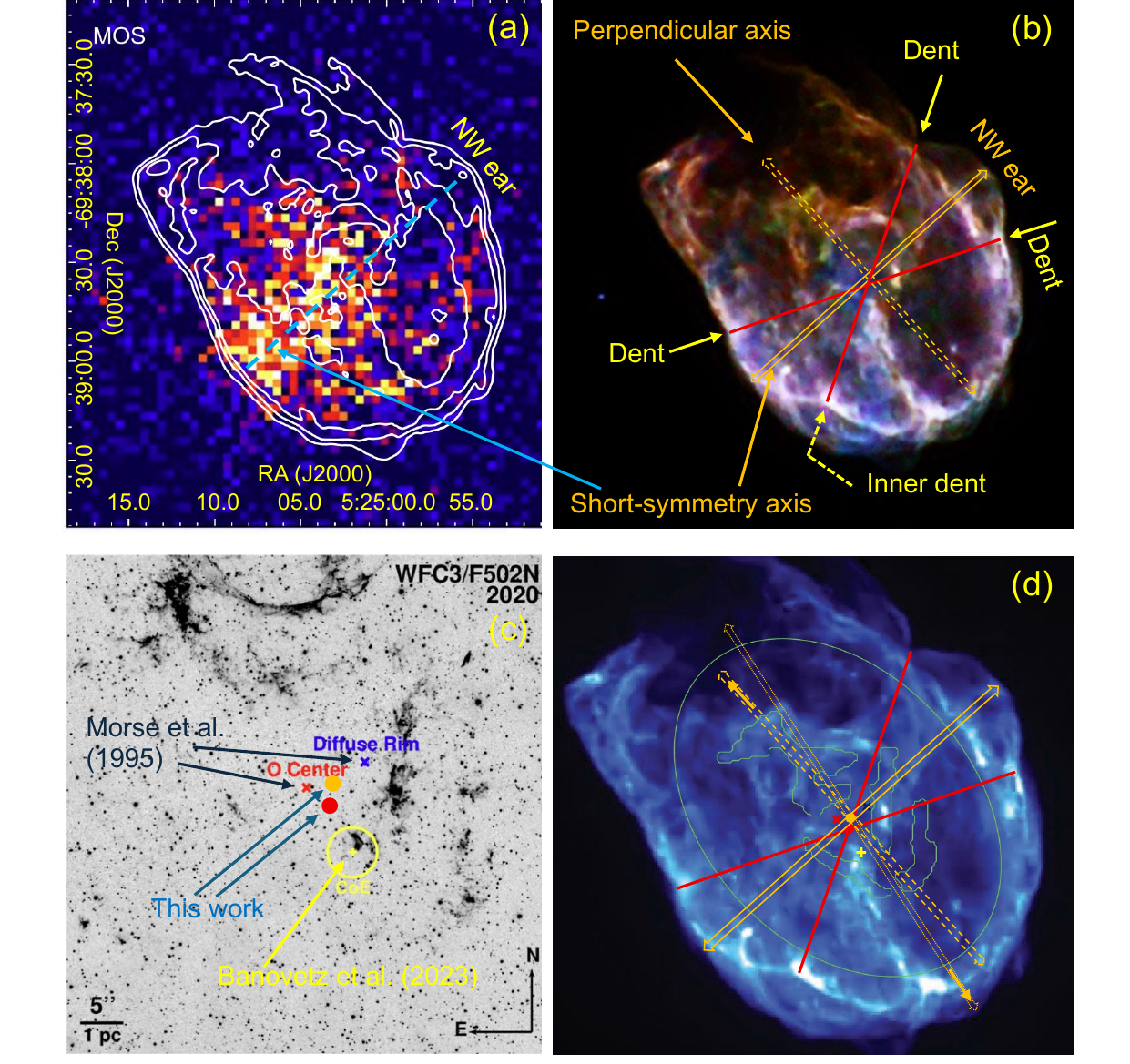} 
\caption{ 
(a) An X-ray image of SNR N132D adapted from \cite{Fosteretal2025}. Colored $2.^{\prime \prime}5 \times 2.^{\prime \prime}5$ squares show the count image in Fe~K$\alpha$ line in the energy range of $6.50-7.05 \keV$, with continuum emission subtracted. The contours are Chandra images in the energy band $0.5-8.0 \keV$. I visually added a line connecting the tip of the northwest ear (protrusion) and the center of the iron distribution. I identify this line as the short-symmetry axis of SNR N132D.    
(b) An X-ray image adapted from the Chandra site (credit: NASA/CXC/NCSU/K.J.Borkowski et al.; {\tiny \url{https://chandra.si.edu/photo/2008/n132d/} }): red for low energy,  green intermediate energy, and blue for high energy emission. I added a double-sided arrow from the tip of the NW ear to the other side at the same location as the dashed, pale blue line in panel (a). The dashed-orange double-sided arrow is perpendicular to the short-symmetry axis arrow and of the same length; they intersect at their centers. I added the two red lines to connect opposite dents. 
(c) An image of oxygen emission adapted from \cite{Banovetzetal2023}, who mark their determination of the center of the proper motion expansion (CoE: yellow plus), and the centers that \cite{Morseetal1995}  found by fitting an ellipse to the diffuse outer rim (blue cross) and the O-rich geometric center (red cross); see Section \ref{subsec:Morse1995}. 
I added the intersection points of the red lines from panel (b) (red dot) and of the two perpendicular double-ended arrows (orange dot). 
(d) A Chandra $0.3-7.0 \keV$ X-ray image of SNR N132D adapted from \cite{Borkowskietal2007}. The size of the image is $120^{\prime \prime} \times 115^{\prime \prime}$, and the scale is $\times 1.25$ that of panels (a) and (b). The four closed green lines near the center of the remnant mark the location of optically emitting O-rich ejecta; these and the ellipse are from the original figure. I copied the two red lines and the two double-sided arrows from panel (b) (increased by a factor of 1.25), as well as the three different SNR centers from panel (c). I added an alternative long-symmetry axis (dotted orange double-sided arrow), which touches the tip of a small ear in the southwest. The intersection of the three double-sided arrows is at the center of each of them. The two solid-orange arrows indicate the possibility that the two opposite jets that shaped the elongation of SNR N132D were at $171^\circ$ to each other (see text).  
}
\label{Fig:SNRN132Dfigure4P}
\end{center}
\end{figure*}

The recent study by \cite{Fosteretal2025} that maps iron emission, as I present in panel (a) of Figure \ref{Fig:SNRN132Dfigure4P}, allows me to identify a symmetry axis, the short-symmetry axis of SNR 
N132D, extending from the northwest ear (NW ear) and through the center of the iron emission; this is the dashed-pale-blue line in panel (a). The X-ray image from the Chandra site, panel (b) of Figure \ref{Fig:SNRN132Dfigure4P}, reveals a clear ear in the northwest. The protrusion on the opposite side is small and not distinct enough to be identified as a separate structural feature, although the two dents are. The new image by \cite{Fosteretal2025} presents a clear opposite structure to the northwest ear: a zone extending from the center to the southeast: the yellow-colored zones in panel (a) of Figure \ref{Fig:SNRN132Dfigure4P} that depict strong Fe K$\alpha$ emission. 
The line along the long axis of the strong Fe K$\alpha$ emission in the southeast crosses to the other side into the northwest ear. This is the short symmetry axis.
The bright zones of other emission lines might be in different regions than the bright Fe K$\alpha$ region, e.g., the S He$\alpha$, Fe He$\alpha$, and Fe Ly$\alpha$ in the XRISM maps \citep{Xrism2024N132D}; these maps have lower spatial resolution than the XMM-Newton maps of \cite{Fosteretal2025}. I use the bright zone of the Fe K$\alpha$ line from \cite{Fosteretal2025} because this zone occupies a well-defined region that is bound by the two dents that I mark on panel (b) of Figure \ref{Fig:SNRN132Dfigure4P}. 

The iron emission Fe~K$\alpha$ is concentrated to the southeast. \cite{Xrism2024N132D} find that the Fe Ly$\alpha$ emission from SNR N132D is redshifted with a bulk velocity of $\simeq 890 \km \s^{-1}$. 
In panel (b) of Figure \ref{Fig:SNRN132Dfigure4P}, I present an X-ray image from the Chandra site. This X-ray image shows three dents (bent inwards) on the outer boundary of the SNR, which I have marked with three yellow arrows. A fourth dent is inside the boundary (yellow-dashed arrow). I connect two pairs of dents with two solid red lines. The solid-orange double-sided arrow is the short-symmetry axis, i.e., at the same location as the dashed line on panel (a) of Figure  \ref{Fig:SNRN132Dfigure4P}. I added a double-sided arrow (dashed orange) for orientation: it is perpendicular to the short-symmetry axis and at the same length; the two double-sided arrows intersect at their centers. As immediately seen, SNR N132D is elongated in the northeast. The southwest end of the perpendicular double-sided arrow does not reach the boundary of the SNR, implying that the SNR is also elongated to the southwest.     

\cite{Banovetzetal2023} measured the proper motion of the oxygen-rich ejecta of SNR N132D from two HST observations 16 years apart. They determined the center of expansion (CoE); in panel (c) of Figure \ref{Fig:SNRN132Dfigure4P}, which I adapted from \cite{Banovetzetal2023}, they mark this center with a yellow plus symbol. They also mark the two centers that \cite{Morseetal1995} identified, which I refer to in Section \ref{subsec:Morse1995}. I added to panel (c) the intersections of the two double-sided arrows and of the two red lines on panel (b) of Figure \ref{Fig:SNRN132Dfigure4P}.  
The Chandra $0.3-7.0 \keV$ X-ray image of SNR N132D in panel (d) of Figure \ref{Fig:SNRN132Dfigure4P}, adapted from \cite{Borkowskietal2007}, shows the global structure of SNR N132D, including the centers of different structures from panel (c), the structures from panel (b), and the contours of the oxygen-rich material near the center (more on the oxygen-rich ejecta in Section \ref{subsec:Morse1995} and \ref{subsec:3D}). 
   
While I based the short-symmetry axis, which is the double-sided arrows from southeast to the NW ear, on structural features, i.e., the NW ear and the iron-rich zone from \cite{Fosteretal2025}, the dashed-orange double-sided arrow is simply a perpendicular line with the same length and through the center of the short-symmetry axis. 
A key feature for identifying point-symmetric morphology is that this perpendicular line aligns with the longest structure of the oxygen-rich ejecta near the center, indicating a clear association. However, I can also draw a line from a small ear on the southwest and through the center of the short-symmetry axis; this is the dotted-orange double-sided arrow. The southwest end of the dotted-orange double-sided arrow touches the edge of that ear, and its center is at the center of the short-symmetry axis. The dashed and dotted orange double-sided arrows are $9^\circ$ to each other. 

I attribute the long extension of SNR N132D to a pair of opposite jets (or more). The angle of $9^\circ$ reflects the uncertainty in the location of the jets' axis. Alternatively, the two jets were not exactly opposite to each other, but rather had an angle of $171^\circ$ between them. Such a `bent symmetry', i.e., the two opposite sides are close to being at $180^\circ$ but not exactly,  is observed in many planetary nebulae. \cite{Shishkinetal2025S147} suggest such bent opposite jets for CCSNR S147. I mark with solid arrows on panel (d) of Figure \ref{Fig:SNRN132Dfigure4P} the two jets I propose for this possibility. 

The main result of this section is the identification of two axes that describe the large-scale structure, the short-symmetry axis and the perpendicular axis that extend along the long dimension of SNR N132D. I turn to show, Sections \ref{subsec:Morse1995} and \ref{subsec:3D}, that these axes correlate with those of the inner structure of the oxygen-rich ejecta. 

\subsection{Previously identified point-symmetric morphology} 
\label{subsec:Morse1995}

\begin{figure}[]
	\begin{center}
\includegraphics[trim=0.0cm 9.5cm 0.0cm 0.0cm ,clip, scale=0.85]{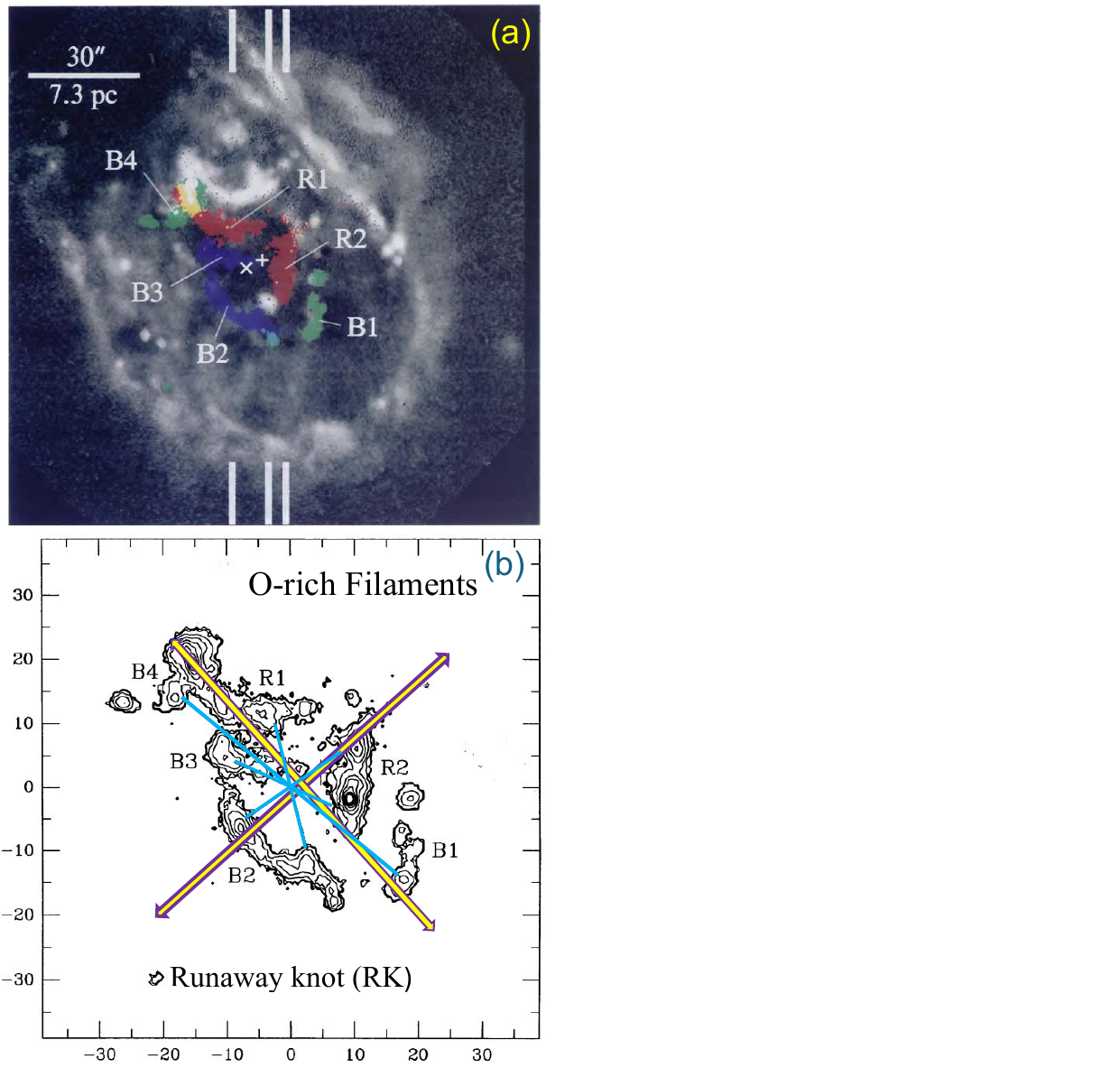} 
\caption{Two panels adapted from \cite{Morseetal1995} and emphasize the oxygen-rich ejecta. 
(a) The velocity map by [O \textsc{iii}]$\lambda$5007 of fast oxygen-rich filaments in SNR N132D on  a gray scale of low-velocity [O \textsc{iii}] emission as presented by \cite{Morseetal1995}: B2 and B3 are highly blueshifted, B1 and B4 are close to the mean velocity, and R1 and R2 are highly redshifted (Section \ref{subsec:3D}). The "+" marks the center of the remnant as \cite{Morseetal1995} determined by fitting an ellipse to the diffuse rim, while the "$\times$" marks the center of the high-velocity oxygen-rich ejecta by the four lines that \cite{Morseetal1995}  mark on panel (b).  
(b) Contours of the oxygen-rich filaments in N132D (panel a). \cite{Morseetal1995} drew the four pale-blue lines by connecting regions which show a symmetric distribution about a common center (the ``$\times$'' symbol on panel a). 
I discuss the runaway knot in Section \ref{subsec:3D}. 
I added the two double-sided arrows from panel (b) of Figure \ref{Fig:SNRN132Dfigure4P}; the arrows are not to scale in length, but only show the directions.   
}
\label{Fig:SNRN132Dfigure2P}
\end{center}
\end{figure}

In Figure \ref{Fig:SNRN132Dfigure2P} I present two panels from \cite{Morseetal1995} of SNR N132D. Panel (a) presents in colors the velocity map by [O \textsc{iii}]$\lambda$5007 of
fast oxygen-rich filaments and in gray the low-velocity oxygen emission. Panel (b) presents the O-rich filaments  with the four lines that \cite{Morseetal1995}
added to the image by connecting regions that
show a symmetric distribution about a common center. They identified a point-symmetric morphological region in the center (although they do not use this term). They identified the intersection of these four lines as the center of the inner, fast, oxygen-rich filament distribution, which they marked by "$\times$" in panel (a). The "$+$" symbol in panel (a) marks the center of the outer diffuse oxygen emission, which they determined by fitting an ellipse to that gas. These two centers also appear in panel (c) of Figure \ref{Fig:SNRN132Dfigure4P}. 

I added to panel (b) of Figure \ref{Fig:SNRN132Dfigure2P} the two double-sided arrows from Figure \ref{Fig:SNRN132Dfigure4P}, that are the short-symmetry axis that I identify, and the perpendicular axis, which is along the long dimension of SNR N132D. From Figures \ref{Fig:SNRN132Dfigure4P} and \ref{Fig:SNRN132Dfigure2P} I notice the following. 
(1) Although the two axes I identify do not coincide with those of \cite{Morseetal1995}, the perpendicular axis is also along the long dimension of the fast oxygen filaments in the center. (2) The center I identify by the two double-sided arrows (orange-dot on panels c and d of Figure \ref{Fig:SNRN132Dfigure4P}) is close to the center of the inner oxygen filaments that \cite{Morseetal1995} identify. 

I conclude that the outer ejecta and the inner ejecta share morphological features. This has significant implications for the shaping mechanism, as the CSM and ISM are unable to shape the inner ejecta. I discuss this and the perpendicular symmetry axes in Section \ref{sec:Summary}. 

\subsection{The relation to the oxygen 3D structure} 
\label{subsec:3D}

\begin{figure}[]
	\begin{center}
\includegraphics[trim=0.0cm 7.5cm 0.0cm 0.0cm ,clip, scale=0.80]{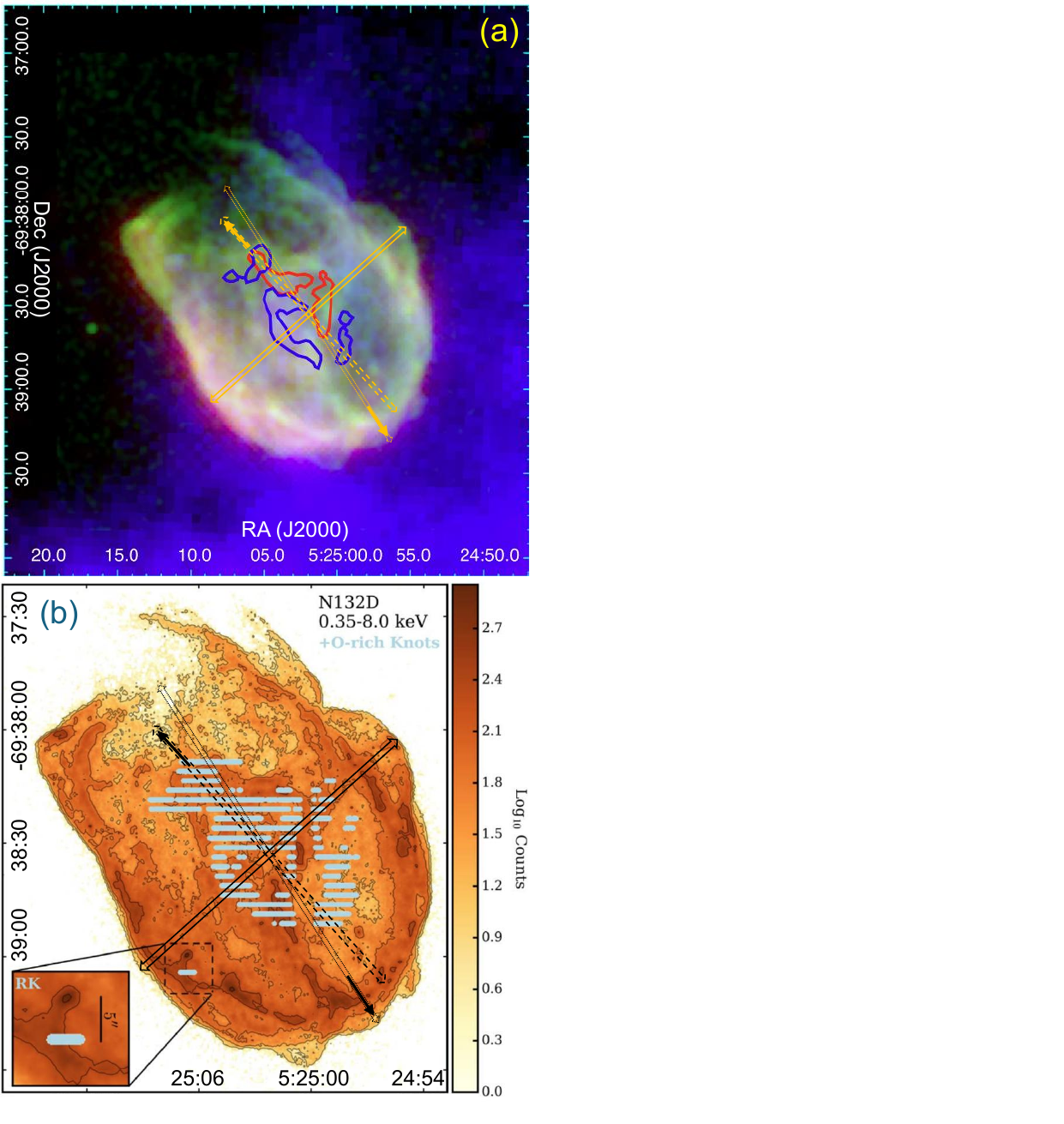} 
\caption{Panels presenting the oxygen-rich torus/ring. To both panels I added the three axes from panel (d) of Figure \ref{Fig:SNRN132Dfigure4P}. 
(a) A multiwavelength three-color image of SNR N132D adapted from \cite{Rhoetal2023}: Herschel $350 \mu$m (blue), Spitzer $24 \mu$m (red), and Chandra X-rays (green). They added the high-velocity blue- and redshifted optical ejecta from \cite{Morseetal1995} as contours of blue and red, respectively. 
(b) An image adapted from \cite{Lawetal2020}, presenting a Chandra image of counts per pixel in the $0.35-8.0 \keV$ band (orange zones), with oxygen-rich optical ejecta overlaid in gray. The inset shows the $5^{\prime \prime} \simeq 1.2 \pc$ offset between the X-ray bright spot and the runaway knot (marked RK on the lower left). 
}
\label{Fig:SNRN132DfigureDynamic}
\end{center}
\end{figure}

\cite{VogtDopita2011} map the [O \textsc{III}]$\lambda$5007 dynamics of SNR N132D and reconstruct its 3D structure. 
They found that the majority of the ejecta form a ring (torus-like) of $\simeq 12 \pc$ in diameter. \cite{Lasker1980} already identified the ring structure of the oxygen-rich ejecta near the center. \cite{VogtDopita2011} speculated that the oxygen-rich ring is in the equatorial plane of a bipolar explosion and that the morphology is also strongly influenced by the CSM. In panel (a) of Figure \ref{Fig:SNRN132DfigureDynamic}, I present an image adapted from \cite{Rhoetal2023} that shows the Doppler shift measurements of \cite{Morseetal1995} on top of the general image of SNR N132D in the IR and X-ray. I added the three axes from panel (d) of Figure \ref{Fig:SNRN132Dfigure4P}. The IR emission (diffuse blue regions) reveals material surrounding SNR N132D; however, I do not attribute the structure of the inner oxygen-rich ring (torus) to the CSM or ISM that are located outside the main SNR shell. 

Additionally, \cite{VogtDopita2011} associated the fast, oxygen-rich runaway knot (RK; panel b of Figure \ref{Fig:SNRN132Dfigure2P}) with a polar jet. I accept this interpretation that polar jets shaped the ring (torus) of oxygen-rich ejecta in the inner part of SNR N132D. I suggest that a pair of opposite jets along the short-symmetry axis participated in the explosion process of SNR N132D; there were several more pairs of jets according to the JJEM. In panel (b) of Figure \ref{Fig:SNRN132DfigureDynamic}, I present the three axes I identified in Figure \ref{Fig:SNRN132Dfigure4P} on top of an image adapted from \cite{Lawetal2020}.
The comparison in the panels of Figure \ref{Fig:SNRN132DfigureDynamic} strengthens the relation I found from Figure \ref{Fig:SNRN132Dfigure2P} between the large-scale symmetry of SNR N132D and the symmetry of the oxygen-rich ejecta in the inner region. 

The 3D structure reconstruction by \cite{Lawetal2020} is broadly consistent with the 3D geometry constructed by \cite{VogtDopita2011}. \cite{Lawetal2020} find the majority of the bright oxygen ejecta to reside in a broken and distorted
torus tilted $\simeq 28^\circ$ to the plane of the sky and with a radius of $4.4 (D/50 \kpc) \pc$, where $D$ is the distance to SNR N132D. The velocity along the line of sight is from $-3000 \km \s^{-1}$ to $+2300 \km \s^{-1}$.
They find the Doppler velocity of the runaway knot to be $\simeq 820 \km \s^{-1}$ and its total space velocity $\simeq 3650 \km \s^{-1}$, about twice the bulk velocity of the oxygen-rich ejecta of $1745 \km \s^{-1}$. Their finding supports the argument of \cite{VogtDopita2011} for a polar jet. 
Of high significance to my claim for the JJEM is \cite{Lawetal2020} finding that the runaway knot is nearly perpendicular to the torus plane and coincident with an X-ray emission spot (inset of panel (b) of Figure \ref{Fig:SNRN132DfigureDynamic}) that is substantially enhanced in silicon and somewhat in sulfur relative to the Large Magellanic Cloud and N132D’s bulk ejecta. 
The different composition from the rest of the ejecta shows that the runaway knot is not a result of an instability, e.g., a finger of Rayleigh-Taylor instability that keeps the same composition stratification of the rest of the ejecta; the runaway knot must be a separate ejection event, namely, a jet. This is the same argument on why the silicon-rich jet of Cassiopeia A cannot be an instability, and it is a jet launched during the explosion process (e.g.,  \citealt{Soker2017IAUS}). 

\cite{Lawetal2020} find some departures from pure axi-symmetry of the torus in that the redshifted material is elevated above the midplane of the torus, while the blueshifted side is below the midplane. In the JJEM, this is part of the point-symmetric morphology that pairs of jets shape during the explosion process.  
\cite{Lawetal2020} also find that the runaway knot is $82^\circ \pm 2$ to the normal to the torus. 

\begin{figure*}[]
	\begin{center}
\includegraphics[trim=0.0cm 16.5cm 0.0cm 0.0cm ,clip, scale=0.85]{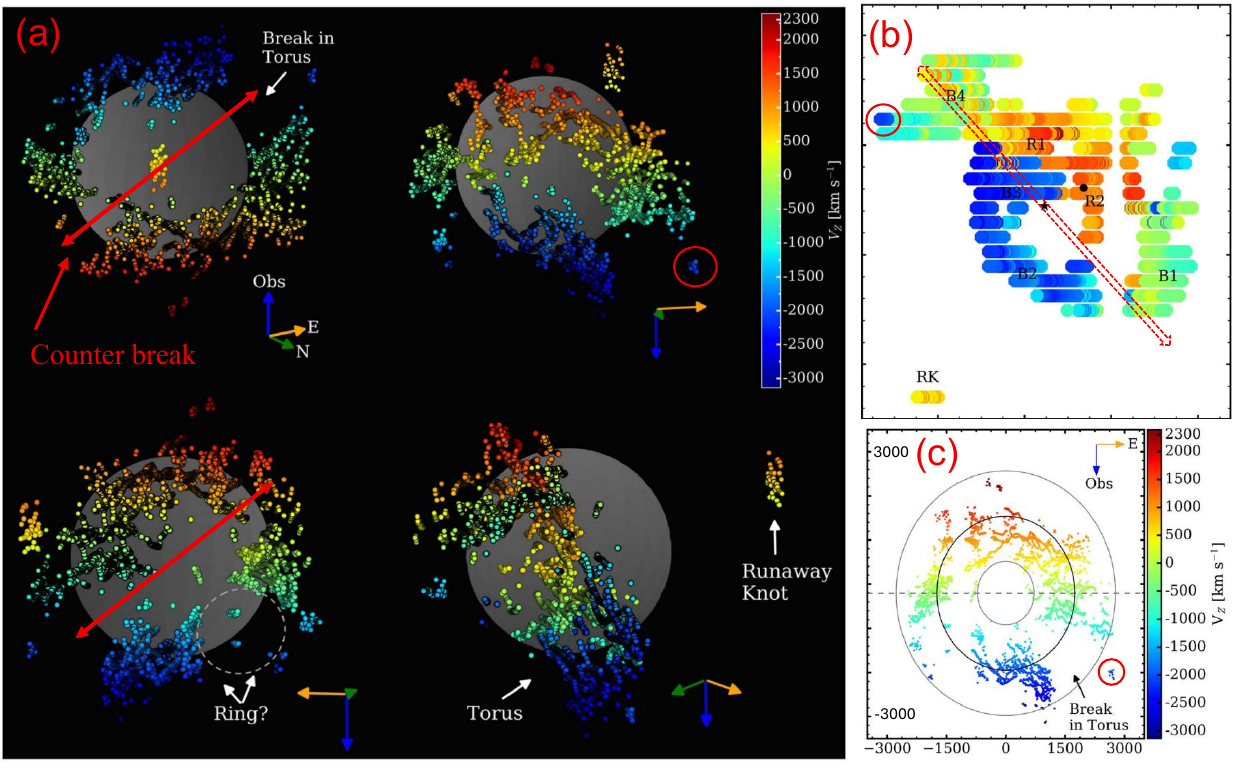} 
\caption{Panels adapted from \cite{Lawetal2020} presenting the optically-emitting oxygen-rich material they analyzed (the gray zones in panel (b) of Figure \ref{Fig:SNRN132DfigureDynamic}). My additions are the marks in red.  
(a) The 3D Doppler reconstructed torus-like structure from different directions. The colors indicate the Doppler shift velocity according to the color bar (from $-3000 \km \s^{-1}$ in blue to $2300 \km \s^{-1}$ in red). The translucent sphere serves as a visual aid to help distinguish between front and back materials.
\cite{Lawetal2020} identified a break in the torus. I identify a counter break that is opposite to the break and narrower in opening. On two panels, I added the double-sided red arrow to connect the break with the counter break. 
(b) Regions with oxygen-rich knots emission, where colors indicate Doppler shift velocity from $-2900 \km \s^{-1}$ in blue to $2500 \km \s^{-1}$ in red. \cite{Lawetal2020} indicated major knots and the two centers from \cite{Morseetal1995}, and the runaway knot (RK). I added the perpendicular axis from Figure \ref{Fig:SNRN132Dfigure4P}, but shifted to go through the center that \cite{Morseetal1995} identified. 
(c) Projections in the plane of the line of sight and the east-west direction of the oxygen-rich material that \cite{Lawetal2020} analyzed. The axes are the expansion velocity in $\km \s^{-1}$. 
In the three panels, I marked the same blue-shifted fast knot with a red circle. 
}
\label{Fig:SNRN132Dfigure3D}
\end{center}
\end{figure*}

\cite{Lawetal2020} identified a break in the torus; I present their identification in Figure \ref{Fig:SNRN132Dfigure3D}. Examining some images from their analysis, I identify a counter break, i.e., a break opposite to the one they identified, as shown in panel (a) of Figure \ref{Fig:SNRN132Dfigure3D}; it has a narrower opening. In the same panel, I draw a line from the break through the center and to the counter break that I identify here. As far as I can tell, this symmetry line through the breaks in the torus coincides on the plane of the sky with the perpendicular axis that I draw in Figures \ref{Fig:SNRN132Dfigure4P} and \ref{Fig:SNRN132Dfigure2P}, up to the uncertainty in the exact location of the center of explosion. I suggest that a jet, one of two jets in a pair, shaped the break in the torus. The continuation of material outward on the sides of the break supports such an interpretation. An opposite jet shaped the counter break. On panel (b) of Figure \ref{Fig:SNRN132Dfigure3D}, I added the perpendicular axis, but shifted to the center that \cite{Lawetal2020} uses. This axis is aligned along the extension of the oxygen-rich material to the northeast, highlighting the relationship between the large-scale symmetry and the oxygen-rich material.    

\cite{Lawetal2020} compare the torus they reconstruct in SNR N132D with that of Cassiopeia A. I note that the torus of Cassiopeia A possesses a rich point-symmetric morphology \citep{BearSoker2025}. I here argue that the torus of SNR N132D also possesses a point-symmetric morphology. \cite{Lawetal2020} compared the runaway knot of SNR N132D to fast ejecta clumps in the Vela SNR, which is another CCSNR with a prominent and rich point-symmetric morphology that only the JJEM can account for (\citealt{Soker2023SNRclass, SokerShishkin2025Vela}). 

The maps of N132D show some structures that are not part of a point-symmetric morphology, like the prominent bright filament extending south–southwest from the center in panels (b) and (d) of Figure \ref{Fig:SNRN132Dfigure4P}. 
This is expected because several processes act to destroy point-symmetry: Instabilities at the explosion process as in the neutrino-driven simulations (see references in Section \ref{sec:intro}), one jet in a pair that is much more energetic than the opposite jet \citep{Soker2024CounterJet, Bearetal2025Puppis}, interaction with an asymmetrical CSM, the NS kick at explosion, nickel bubbles which are concentrations of nickel that when decay heat the gas and inflate bubbles, and binary star at explosion (see discussion by \citealt{SokerShishkin2025Vela}). This specific filament seems to result from unequal jets with instabilities that increased the asymmetry. This is for the future three-dimensional simulations of the JJEM. 

This section solidifies my identification of a point-symmetric morphology in SNR N132D, and its relation to some properties in other point-symmetric CCSNRs. 

\section{Discussion and Summary} 
\label{sec:Summary}

By visually inspecting images of CCSNR N132D, a method I justified in Section \ref{sec:PointSymmetry}, I identified a point-symmetric morphology in CCSNR N132D, composed of two symmetry axes. The first one is the short symmetry axis, extending from the northwest ear and through the center of the iron-rich emission that \cite{Fosteretal2025} presented in a recent study (panel a of Figure \ref{Fig:SNRN132Dfigure4P}). The second, the perpendicular axis on the plane of the sky, is an axis perpendicular to the short symmetry axis and passing through its center, which is also aligned along the long dimension of SNR N132D; it coincides with the extension of the oxygen-rich gas to the northeast (panel d of Figure \ref{Fig:SNRN132Dfigure4P}, Figure \ref{Fig:SNRN132Dfigure2P} and \ref{Fig:SNRN132DfigureDynamic}, and panel b of Figure \ref{Fig:SNRN132Dfigure3D}). The perpendicular axis seems to coincide with the line from the break in the torus that \cite{Lawetal2020} identify to the counter break that I identify (panel a of Figure \ref{Fig:SNRN132Dfigure3D}). There might be a third axis: \cite{Lawetal2020} find the redshifted material in the torus to be somewhat above the midplane of the torus and the blueshifted side to be below the midplane. A line between these two regions, that is inclined to the plane of the torus, might define a third symmetry axis in SNR N132D. This deserves further study. 

SNR N132D is surrounded by material (CSM and/or ISM), as indicated by the diffuse blue zones in panel (a) of Figure \ref{Fig:SNRN132DfigureDynamic}. SNR N132D interacts with the surrounding clouds (e.g., \citealt{Dopitaetal2018, Sanoetal2020, HESSColl2021, Guetal2025N132D}). This interaction affects the morphology of the outer regions, but cannot shape the inner oxygen-rich gas. The fact that the inner and outer morphologies correlate shows that the point symmetry is a result of the explosion. 

I consider two energetic pairs of jets to shape the two symmetry axes; other, weaker pairs of jets that contributed to the explosion process are possible and likely. 
One pair of energetic jets is perpendicular to the plane of the oxygen-rich torus (Figure \ref{Fig:SNRN132Dfigure3D}). This jet axis coincides with the short symmetry axis. The runaway knot (Figures \ref{Fig:SNRN132Dfigure2P} - \ref{Fig:SNRN132Dfigure3D}) is a remnant of one of these two energetic jets. The unequal sides, an ear on the northwest and an iron-rich zone in the southeast (panel a of Figure \ref{Fig:SNRN132Dfigure4P}), suggest that the jets were unequal in their power and/or opening angles. This deserves further study. The second pair of jets is perpendicular to the first pair of jets and is aligned along the long dimension of SNR N132D, coinciding with the long axis of the oxygen-rich material, as shown in all figures. The two jets were unequal, with the northeast jet being significantly more powerful, as indicated by the more extended SNR in that direction.  

I considered the possibility that the two opposite jets along the long dimension of SNR N132D were not exactly opposite, i.e., a bent pair of jets. I mark the two suggested jets' directions with solid-orange arrows in panel (d) of Figure \ref{Fig:SNRN132Dfigure4P}; they are at $171^\circ$ to each other. \cite{Shishkinetal2025S147} suggested that one of the pairs of jets that exploded and shaped CCSNR S147 was a bent pair. 

The JJEM accounts for the point-symmetric morphology of SNR N132D, as well as for the perpendicular two axes and the bent pair of jets. The perpendicular axes might result from the following two effects. In an early energetic pair of jets, one jet carries much more momentum (and energy) than the other. As a result of that, the NS acquires a kick velocity; this is the kick-BEAP (kick by early asymmetrical pair) mechanism \citep{Bearetal2025Puppis, Shishkinetal2025S147}. The kick velocity is along the axis of the jets. The kick velocity imparts an angular momentum component to the material that the NS accretes later from the collapsing core that is perpendicular to the kick velocity. If this component dominates, then the pair of jets that the accreted gas launches along its angular momentum axis tends to be perpendicular to the kick velocity (e.g., \citealt{BearSoker2018kick}), i.e., perpendicular to the axis of the first pair of jets. This effect of post-kick accretion might explain the avoidance of small angles between the NS kick velocity and the main jet axis in some CCSNRs (e.g., \citealt{Soker2022SNR0540, BearSoker2023RNAAS}). 
Many of the accretion episodes through the intermittent accretion disks in the JJEM last for a time scale that is shorter, or not much longer than the relaxation time of the accretion disk \citep{Soker2024CounterJet}. Because accretion of material with stochastic angular momentum and density fluctuations forms the accretion disk, the two sides of the unrelaxed accretion disk might be unequal in size and structure. Earlier studies of the JJEM followed \citep{Soker2024CounterJet} and considered that opposite jets differ in their power and opening angle. I here add to the claim by \cite{Shishkinetal2025S147} that, in addition, the two jets might not be exactly opposite to each other, and the angle between them can be $<180^\circ$, e.g., $\simeq 150^\circ$ in SNR S147. 

One of the future steps will be to conduct heavy three-dimensional hydrodynamical simulations of the expansion of the jets through the inner core, i.e., in the silicon and oxygen layers, and follow the nucleosynthesis that the jets induce. These simulations will find the kinematics of the different elements. They will account for the dynamics, e.g., the finding that the velocity dispersions of the sulfur and silicon emission lines in SNR N132D are $\simeq 450 \km \s^{-1}$ while that of the Fe He$\alpha$ emission line is $\simeq 1700 \km \s^{-1}$ \citep{Xrism2024N132D} 

SNR N132D is the fifteenth CCSNR with an identified point-symmetric morphology attributed to the JJEM (Section \ref{sec:PointSymmetry}). Presently, the point-symmetric morphology is the only property of CCSNe and CCSNRs that can clearly distinguish between the JJEM and the delayed neutrino explosion mechanism \citep{Soker2024UnivReview}. Other properties, such as light curves and nucleosynthesis, are similar but not identical between the two explosion mechanisms. However, the differences in these properties are not at a level that allows for a clear observational determination of the explosion mechanism \citep{Soker2024UnivReview}. The point-symmetrical morphology due to point-symmetric explosion is a robust prediction of the JJEM (although it will not be observed in all CCSNRs) that has no explanation in the neutrino-driven mechanism (e.g., \citealt{SokerShishkin2025Vela}). 
Some supporters of the neutrino-driven explosion mechanism have noted that their simulations do not produce jets (I attributed it to the limited ability of existing CCSN hydrodynamical numerical codes to handle magnetic field reconnection; \citealt{Soker2025Learning}). As a result, they often do not engage with or cite literature related to the JJEM model. This approach, where theoretical expectations (but only by some researchers) are used to dismiss observational evidence, contrasts with the standard scientific methodology in which models are tested and refined based on empirical data. In contrast, the JJEM model argues that the observed point symmetric morphologies in CCSNRs provide significant evidence that favors jet-driven processes over neutrino-driven explosions as the primary explosion mechanism. I therefore encourage the CCSNe research community to consider evaluating the JJEM on equal ground with the delayed neutrino explosion mechanism.

\section*{Acknowledgements}

I thank Ealeal Bear, Dima Shishkin, and Amit Kashi for their many discussions and comments, which were relevant to the topic of this study. I thank an anonymous referee for useful comments.  I thank the Charles Wolfson Academic Chair at the Technion for the support.


 \bibliography{reference}{}

@ARTICLE{BearSoker2023RNAAS,
       author = {{Bear}, Ealeal and {Soker}, Noam},
        title = "{The Jets and the Neutron Star Kick Velocity of the Supernova Remnant CTB 1}",
      journal = {Research Notes of the American Astronomical Society},
         year = 2023,
        month = dec,
       volume = {7},
       number = {12},
          eid = {266},
        pages = {266},
          doi = {10.3847/2515-5172/ad1392},
archivePrefix = {arXiv},
       eprint = {2312.02026},
 primaryClass = {astro-ph.HE},
       adsurl = {https://ui.adsabs.harvard.edu/abs/2023RNAAS...7..266B}
}

@ARTICLE{GilkisSoker2014,
       author = {{Gilkis}, Avishai and {Soker}, Noam},
        title = "{Triggering jet-driven explosions of core-collapse supernovae by accretion from convective regions}",
      journal = {\mnras},
         year = 2014,
        month = apr,
       volume = {439},
       number = {4},
        pages = {4011-4017},
          doi = {10.1093/mnras/stu257},
archivePrefix = {arXiv},
       eprint = {1401.1597},
 primaryClass = {astro-ph.SR},
       adsurl = {https://ui.adsabs.harvard.edu/abs/2014MNRAS.439.4011G}
}

@ARTICLE{GilkisSoker2016,
       author = {{Gilkis}, Avishai and {Soker}, Noam},
        title = "{Angular Momentum Fluctuations in the Convective Helium Shell of Massive Stars}",
      journal = {\apj},
         year = 2016,
        month = aug,
       volume = {827},
       number = {1},
          eid = {40},
        pages = {40},
          doi = {10.3847/0004-637X/827/1/40},
archivePrefix = {arXiv},
       eprint = {1505.05756},
 primaryClass = {astro-ph.SR},
       adsurl = {https://ui.adsabs.harvard.edu/abs/2016ApJ...827...40G}
}

@ARTICLE{Guettaetal2020,
       author = {{Guetta}, Dafne and {Rahin}, Roi and {Bartos}, Imre and {Della Valle}, Massimo},
        title = "{Constraining the fraction of core-collapse supernovae harbouring choked jets with high-energy neutrinos}",
      journal = {\mnras},
         year = 2020,
        month = feb,
       volume = {492},
       number = {1},
        pages = {843-847},
          doi = {10.1093/mnras/stz3245},
archivePrefix = {arXiv},
       eprint = {1906.07399},
 primaryClass = {astro-ph.HE},
       adsurl = {https://ui.adsabs.harvard.edu/abs/2020MNRAS.492..843G}
}

@ARTICLE{Hughes1987,
       author = {{Hughes}, John P.},
        title = "{X-Ray Studies of the Supernova Remnant N132D. I. Morphology}",
      journal = {\apj},
         year = 1987,
        month = mar,
       volume = {314},
        pages = {103},
          doi = {10.1086/165043},
       adsurl = {https://ui.adsabs.harvard.edu/abs/1987ApJ...314..103H}
}

@ARTICLE{PapishSoker2011,
       author = {{Papish}, Oded and {Soker}, Noam},
        title = "{Exploding core collapse supernovae with jittering jets}",
      journal = {\mnras},
         year = 2011,
        month = sep,
       volume = {416},
       number = {3},
        pages = {1697-1702},
          doi = {10.1111/j.1365-2966.2011.18671.x},
archivePrefix = {arXiv},
       eprint = {1103.1554},
 primaryClass = {astro-ph.HE},
       adsurl = {https://ui.adsabs.harvard.edu/abs/2011MNRAS.416.1697P}
}

@ARTICLE{ShishkinSoker2021,
       author = {{Shishkin}, Dmitry and {Soker}, Noam},
        title = "{Supplying angular momentum to the jittering jets explosion mechanism using inner convection layers}",
      journal = {\mnras},
         year = 2021,
        month = nov,
       volume = {508},
       number = {1},
        pages = {L43-L47},
          doi = {10.1093/mnrasl/slab105},
archivePrefix = {arXiv},
       eprint = {2107.08779},
 primaryClass = {astro-ph.HE},
       adsurl = {https://ui.adsabs.harvard.edu/abs/2021MNRAS.508L..43S}
}

@ARTICLE{ShishkinSoker2023,
       author = {{Shishkin}, Dmitry and {Soker}, Noam},
        title = "{The implications of large binding energies of massive stripped core collapse supernova progenitors on the explosion mechanism}",
      journal = {\mnras},
         year = 2023,
        month = jun,
       volume = {522},
       number = {1},
        pages = {438-445},
          doi = {10.1093/mnras/stad889},
archivePrefix = {arXiv},
       eprint = {2301.05144},
 primaryClass = {astro-ph.HE},
       adsurl = {https://ui.adsabs.harvard.edu/abs/2023MNRAS.522..438S}
}

@ARTICLE{Soker2010,
       author = {{Soker}, Noam},
        title = "{Applying the jet feedback mechanism to core-collapse supernova explosions}",
      journal = {\mnras},
         year = 2010,
        month = feb,
       volume = {401},
       number = {4},
        pages = {2793-2798},
          doi = {10.1111/j.1365-2966.2009.15862.x},
archivePrefix = {arXiv},
       eprint = {0909.5276},
 primaryClass = {astro-ph.HE},
       adsurl = {https://ui.adsabs.harvard.edu/abs/2010MNRAS.401.2793S}
}

@ARTICLE{Soker2022SNR0540,
       author = {{Soker}, Noam},
        title = "{Imprints of the Jittering Jets Explosion Mechanism in the Morphology of the Supernova Remnant SNR 0540-69.3}",
      journal = {Research in Astronomy and Astrophysics},
         year = 2022,
        month = mar,
       volume = {22},
       number = {3},
          eid = {035019},
        pages = {035019},
          doi = {10.1088/1674-4527/ac49e6},
archivePrefix = {arXiv},
       eprint = {2109.10230},
 primaryClass = {astro-ph.HE},
       adsurl = {https://ui.adsabs.harvard.edu/abs/2022RAA....22c5019S}
}

@ARTICLE{Soker2023SNRclass,
       author = {{Soker}, Noam},
        title = "{Classifying Core Collapse Supernova Remnants by Their Morphology as Shaped by the Last Exploding Jets}",
      journal = {Research in Astronomy and Astrophysics},
         year = 2023,
        month = nov,
       volume = {23},
       number = {11},
          eid = {115017},
        pages = {115017},
          doi = {10.1088/1674-4527/acf446},
archivePrefix = {arXiv},
       eprint = {2307.15666},
 primaryClass = {astro-ph.HE},
       adsurl = {https://ui.adsabs.harvard.edu/abs/2023RAA....23k5017S}
}

@ARTICLE{Soker2024NA1987A,
       author = {{Soker}, Noam},
        title = "{Hints of point-symmetric structures in SN 1987A: The jittering jets explosion mechanism}",
      journal = {\na},
         year = 2024,
        month = apr,
       volume = {107},
          eid = {102154},
        pages = {102154},
          doi = {10.1016/j.newast.2023.102154},
archivePrefix = {arXiv},
       eprint = {2309.07863},
 primaryClass = {astro-ph.HE},
       adsurl = {https://ui.adsabs.harvard.edu/abs/2024NewA..10702154S}
}

@ARTICLE{WangShishkinSoker2024,
       author = {{Wang}, Nikki Yat Ning and {Shishkin}, Dmitry and {Soker}, Noam},
        title = "{The Jittering Jets Explosion Mechanism in Electron Capture Supernovae}",
      journal = {\apj},
         year = 2024,
        month = jul,
       volume = {969},
       number = {2},
          eid = {163},
        pages = {163},
          doi = {10.3847/1538-4357/ad487f},
archivePrefix = {arXiv},
       eprint = {2401.06652},
 primaryClass = {astro-ph.HE},
       adsurl = {https://ui.adsabs.harvard.edu/abs/2024ApJ...969..163W}
}

@ARTICLE{Soker2024CF,
       author = {{Soker}, Noam},
        title = "{Comparing jet-shaped point symmetry in cluster cooling flows and supernovae}",
      journal = {The Open Journal of Astrophysics},
         year = 2024,
        month = jun,
       volume = {7},
          eid = {49},
        pages = {49},
          doi = {10.33232/001c.120279},
archivePrefix = {arXiv},
       eprint = {2403.08544},
 primaryClass = {astro-ph.HE},
       adsurl = {https://ui.adsabs.harvard.edu/abs/2024OJAp....7...49S}
}

@ARTICLE{Soker2024CounterJet,
       author = {{Soker}, Noam},
        title = "{Jet - counter-jet asymmetry in the jittering jets explosion mechanism of supernovae}",
      journal = {The Open Journal of Astrophysics},
         year = 2024,
        month = feb,
       volume = {7},
          eid = {12},
        pages = {12},
          doi = {10.21105/astro.2311.03286},
archivePrefix = {arXiv},
       eprint = {2311.03286},
 primaryClass = {astro-ph.HE},
       adsurl = {https://ui.adsabs.harvard.edu/abs/2024OJAp....7E..12S}
}

@ARTICLE{Velazquezetal2023,
       author = {{Vel{\'a}zquez}, P.~F. and {Meyer}, D.~M. -A. and {Chiotellis}, A. and {Cruz-{\'A}lvarez}, A.~E. and {Schneiter}, E.~M. and {Toledo-Roy}, J.~C. and {Reynoso}, E.~M. and {Esquivel}, A.},
        title = "{The sculpting of rectangular and jet-like morphologies in supernova remnants by anisotropic equatorially confined progenitor stellar winds}",
      journal = {\mnras},
         year = 2023,
        month = mar,
       volume = {519},
       number = {4},
        pages = {5358-5372},
          doi = {10.1093/mnras/stad039},
archivePrefix = {arXiv},
       eprint = {2301.03660},
 primaryClass = {astro-ph.HE},
       adsurl = {https://ui.adsabs.harvard.edu/abs/2023MNRAS.519.5358V}
}

@ARTICLE{Burrowsetal2024kick,
       author = {{Burrows}, Adam and {Wang}, Tianshu and {Vartanyan}, David and {Coleman}, Matthew S.~B.},
        title = "{A Theory for Neutron Star and Black Hole Kicks and Induced Spins}",
      journal = {\apj},
         year = 2024,
        month = mar,
       volume = {963},
       number = {1},
          eid = {63},
        pages = {63},
          doi = {10.3847/1538-4357/ad2353},
       adsurl = {https://ui.adsabs.harvard.edu/abs/2024ApJ...963...63B}
}

@ARTICLE{ZhaMullerPowell2024,
       author = {{Zha}, Shuai and {M{\"u}ller}, Bernhard and {Powell}, Jade},
        title = "{Nucleosynthesis in the Innermost Ejecta of Magnetorotational Supernova Explosions in 3-dimensions}",
      journal = {arXiv e-prints},
         year = 2024,
        month = mar,
          eid = {arXiv:2403.02072},
        pages = {arXiv:2403.02072},
          doi = {10.48550/arXiv.2403.02072},
archivePrefix = {arXiv},
       eprint = {2403.02072},
 primaryClass = {astro-ph.HE},
       adsurl = {https://ui.adsabs.harvard.edu/abs/2024arXiv240302072Z}
}

@ARTICLE{vanBaaletal2024,
       author = {{van Baal}, Bart F.~A. and {Jerkstrand}, Anders and {Wongwathanarat}, Annop and {Janka}, Hans-Thomas},
        title = "{Diagnostics of 3D explosion asymmetries of stripped-envelope supernovae by nebular line profiles}",
      journal = {\mnras},
         year = 2024,
        month = jun,
          doi = {10.1093/mnras/stae1603},
archivePrefix = {arXiv},
       eprint = {2404.01763},
 primaryClass = {astro-ph.HE},
       adsurl = {https://ui.adsabs.harvard.edu/abs/2024MNRAS.tmp.1575V}
}

@ARTICLE{WangBurrows2024,
       author = {{Wang}, Tianshu and {Burrows}, Adam},
        title = "{Supernova Explosions of the Lowest-mass Massive Star Progenitors}",
      journal = {\apj},
         year = 2024,
        month = jul,
       volume = {969},
       number = {2},
          eid = {74},
        pages = {74},
          doi = {10.3847/1538-4357/ad5009},
archivePrefix = {arXiv},
       eprint = {2405.06024},
 primaryClass = {astro-ph.HE},
       adsurl = {https://ui.adsabs.harvard.edu/abs/2024ApJ...969...74W}
}

@ARTICLE{ Shibagakietal2024,
       author = {{Shibagaki}, Shota and {Kuroda}, Takami and {Kotake}, Kei and {Takiwaki}, Tomoya and {Fischer}, Tobias},
        title = "{Three-dimensional GRMHD simulations of rapidly rotating stellar core collapse}",
      journal = {\mnras},
         year = 2024,
        month = jul,
       volume = {531},
       number = {3},
        pages = {3732-3743},
          doi = {10.1093/mnras/stae1361},
archivePrefix = {arXiv},
       eprint = {2309.05161},
 primaryClass = {astro-ph.HE},
       adsurl = {https://ui.adsabs.harvard.edu/abs/2024MNRAS.531.3732S}
}

@ARTICLE{JankaKresse2024,
       author = {{Janka}, H. -Thomas and {Kresse}, Daniel},
        title = "{Interplay Between Neutrino Kicks and Hydrodynamic Kicks of Neutron Stars and Black Holes}",
      journal = {arXiv e-prints},
         year = 2024,
        month = jan,
          eid = {arXiv:2401.13817},
        pages = {arXiv:2401.13817},
          doi = {10.48550/arXiv.2401.13817},
archivePrefix = {arXiv},
       eprint = {2401.13817},
 primaryClass = {astro-ph.HE},
       adsurl = {https://ui.adsabs.harvard.edu/abs/2024arXiv240113817J}
}

@ARTICLE{Soker2022nu,
       author = {{Soker}, Noam},
        title = "{Boosting Jittering Jets by Neutrino Heating in Core Collapse Supernovae}",
      journal = {Research in Astronomy and Astrophysics},
         year = 2022,
        month = sep,
       volume = {22},
       number = {9},
          eid = {095007},
        pages = {095007},
          doi = {10.1088/1674-4527/ac7cbc},
archivePrefix = {arXiv},
       eprint = {2202.05556},
 primaryClass = {astro-ph.HE},
       adsurl = {https://ui.adsabs.harvard.edu/abs/2022RAA....22i5007S}
}

@ARTICLE{Nakamuraetal2025,
       author = {{Nakamura}, Ko and {Takiwaki}, Tomoya and {Matsumoto}, Jin and {Kotake}, Kei},
        title = "{Three-dimensional magnetohydrodynamic simulations of core-collapse supernovae - I. Hydrodynamic evolution and protoneutron star properties}",
      journal = {\mnras},
         year = 2025,
        month = jan,
       volume = {536},
       number = {1},
        pages = {280-294},
          doi = {10.1093/mnras/stae2611},
archivePrefix = {arXiv},
       eprint = {2405.08367},
 primaryClass = {astro-ph.HE},
       adsurl = {https://ui.adsabs.harvard.edu/abs/2025MNRAS.536..280N}
}

@ARTICLE{Andresenetal2024,
       author = {{Andresen}, Haakon and {O'Connor}, Evan P. and {Andersen}, Oliver Eggenberger and {Couch}, Sean M.},
        title = "{Gray two-moment neutrino transport: Comprehensive tests and improvements for supernova simulations}",
      journal = {\aap},
         year = 2024,
        month = jul,
       volume = {687},
          eid = {A55},
        pages = {A55},
          doi = {10.1051/0004-6361/202449776},
archivePrefix = {arXiv},
       eprint = {2402.18303},
 primaryClass = {astro-ph.HE},
       adsurl = {https://ui.adsabs.harvard.edu/abs/2024A&A...687A..55A}
}

@ARTICLE{MeyerMelianietal2024,
       author = {{Meyer}, D.~M. -A. and {Meliani}, Z. and {Vel{\'a}zquez}, P.~F. and {Pohl}, M. and {Torres}, D.~F.},
        title = "{On the plerionic rectangular supernova remnants of static progenitors}",
      journal = {\mnras},
         year = 2024,
        month = jan,
       volume = {527},
       number = {3},
        pages = {5514-5524},
          doi = {10.1093/mnras/stad3495},
archivePrefix = {arXiv},
       eprint = {2311.06817},
 primaryClass = {astro-ph.HE},
       adsurl = {https://ui.adsabs.harvard.edu/abs/2024MNRAS.527.5514M}
}

@ARTICLE{MeyerDetal2024,
       author = {{Meyer}, D.~M. -A. and {Vel{\'a}zquez}, P.~F. and {Pohl}, M. and {Egberts}, K. and {Petrov}, M. and {Villagran}, M.~A. and {Torres}, D.~F. and {Batzofin}, R.},
        title = "{Supernova remnants of red supergiants: From barrels to loops}",
      journal = {\aap},
         year = 2024,
        month = jul,
       volume = {687},
          eid = {A127},
        pages = {A127},
          doi = {10.1051/0004-6361/202449706},
       adsurl = {https://ui.adsabs.harvard.edu/abs/2024A&A...687A.127M}
}

@ARTICLE{Meyeretal2022,
       author = {{Meyer}, D.~M. -A. and {Vel{\'a}zquez}, P.~F. and {Petruk}, O. and {Chiotellis}, A. and {Pohl}, M. and {Camps-Fari{\~n}a}, A. and {Petrov}, M. and {Reynoso}, E.~M. and {Toledo-Roy}, J.~C. and {Schneiter}, E.~M. and {Castellanos-Ram{\'\i}rez}, A. and {Esquivel}, A.},
        title = "{Rectangular core-collapse supernova remnants: application to Puppis A}",
      journal = {\mnras},
         year = 2022,
        month = sep,
       volume = {515},
       number = {1},
        pages = {594-605},
          doi = {10.1093/mnras/stac1832},
archivePrefix = {arXiv},
       eprint = {2206.14495},
 primaryClass = {astro-ph.HE},
       adsurl = {https://ui.adsabs.harvard.edu/abs/2022MNRAS.515..594M}
}

@ARTICLE{ChiotellisZapartasMeyer2024,
       author = {{Chiotellis}, Alexandros and {Zapartas}, Emmanouil and {Meyer}, Dominique M. -A.},
        title = "{On the origin of mixed morphology supernova remnants: linking their properties to the evolution of a red supergiant progenitor star}",
      journal = {\mnras},
         year = 2024,
        month = jul,
       volume = {531},
       number = {4},
        pages = {5109-5116},
          doi = {10.1093/mnras/stae947},
archivePrefix = {arXiv},
       eprint = {2403.19743},
 primaryClass = {astro-ph.HE},
       adsurl = {https://ui.adsabs.harvard.edu/abs/2024MNRAS.531.5109C}
}

@ARTICLE{LuYanetal2021,
       author = {{Lu}, Chun-Yan and {Yan}, Jing-Wen and {Wen}, Lu and {Fang}, Jun},
        title = "{Numerically investigating the peculiar periphery of a supernova remnant in the medium with a density gradient: the case of RCW 103}",
      journal = {Research in Astronomy and Astrophysics},
         year = 2021,
        month = mar,
       volume = {21},
       number = {2},
          eid = {033},
        pages = {033},
          doi = {10.1088/1674-4527/21/2/33},
archivePrefix = {arXiv},
       eprint = {2008.02574},
 primaryClass = {astro-ph.HE},
       adsurl = {https://ui.adsabs.harvard.edu/abs/2021RAA....21...33L}
}

@ARTICLE{YanLuetal2020,
       author = {{Yan}, Jing-Wen and {Lu}, Chun-Yan and {Wen}, Lu and {Yu}, Huan and {Fang}, Jun},
        title = "{Investigating the morphology of the supernova remnant G349.7+00.2 in the medium with a density gradient}",
      journal = {Research in Astronomy and Astrophysics},
         year = 2020,
        month = sep,
       volume = {20},
       number = {9},
          eid = {154},
        pages = {154},
          doi = {10.1088/1674-4527/20/9/154},
archivePrefix = {arXiv},
       eprint = {2004.06992},
 primaryClass = {astro-ph.HE},
       adsurl = {https://ui.adsabs.harvard.edu/abs/2020RAA....20..154Y}
}

@ARTICLE{Wuetal2019,
       author = {{Wu}, Dan and {Zhang}, Meng-Fei},
        title = "{How does a strong surrounding magnetic field influence the evolution of a supernova remnant?}",
      journal = {Research in Astronomy and Astrophysics},
         year = 2019,
        month = sep,
       volume = {19},
       number = {9},
          eid = {124},
        pages = {124},
          doi = {10.1088/1674-4527/19/9/124},
       adsurl = {https://ui.adsabs.harvard.edu/abs/2019RAA....19..124W}
}

@ARTICLE{Chiotellisetal2021,
       author = {{Chiotellis}, A. and {Boumis}, P. and {Spetsieri}, Z.~T.},
        title = "{'Ears' formation in supernova remnants: overhearing an interaction history with bipolar circumstellar structures}",
      journal = {\mnras},
         year = 2021,
        month = mar,
       volume = {502},
       number = {1},
        pages = {176-187},
          doi = {10.1093/mnras/staa3573},
archivePrefix = {arXiv},
       eprint = {2011.06020},
 primaryClass = {astro-ph.SR},
       adsurl = {https://ui.adsabs.harvard.edu/abs/2021MNRAS.502..176C}
}

@ARTICLE{Buelletetal2023,
       author = {{Buellet}, A. -C. and {Foglizzo}, T. and {Guilet}, J. and {Abdikamalov}, E.},
        title = "{Effect of stellar rotation on the development of post-shock instabilities during core-collapse supernovae}",
      journal = {\aap},
         year = 2023,
        month = jun,
       volume = {674},
          eid = {A205},
        pages = {A205},
          doi = {10.1051/0004-6361/202245799},
archivePrefix = {arXiv},
       eprint = {2301.01962},
 primaryClass = {astro-ph.HE},
       adsurl = {https://ui.adsabs.harvard.edu/abs/2023A&A...674A.205B}
}

@ARTICLE{Soker2024Keyhole,
       author = {{Soker}, Noam},
        title = "{Supernova 1987A's Keyhole: A Long-lived Jet-pair in the Final Explosion Phase of Core-collapse Supernovae}",
      journal = {Research in Astronomy and Astrophysics},
         year = 2024,
        month = jul,
       volume = {24},
       number = {7},
          eid = {075006},
        pages = {075006},
          doi = {10.1088/1674-4527/ad4fc2},
archivePrefix = {arXiv},
       eprint = {2404.07455},
 primaryClass = {astro-ph.HE},
       adsurl = {https://ui.adsabs.harvard.edu/abs/2024RAA....24g5006S}
}

@ARTICLE{BearSoker2018kick,
       author = {{Bear}, Ealeal and {Soker}, Noam},
        title = "{Neutron Star Natal Kick and Jets in Core Collapse Supernovae}",
      journal = {\apj},
         year = 2018,
        month = mar,
       volume = {855},
       number = {2},
          eid = {82},
        pages = {82},
          doi = {10.3847/1538-4357/aaad07},
archivePrefix = {arXiv},
       eprint = {1710.00819},
 primaryClass = {astro-ph.HE},
       adsurl = {https://ui.adsabs.harvard.edu/abs/2018ApJ...855...82B}
}

@ARTICLE{ShishkinKayeSoker2024,
       author = {{Shishkin}, Dmitry and {Kaye}, Roy and {Soker}, Noam},
        title = "{Identifying jittering-jet-shaped ejecta in the Cygnus Loop supernova remnant}",
      journal = {arXiv e-prints},
         year = 2024,
        month = aug,
          eid = {arXiv:2408.11014},
        pages = {arXiv:2408.11014},
archivePrefix = {arXiv},
       eprint = {2408.11014},
 primaryClass = {astro-ph.HE},
       adsurl = {https://ui.adsabs.harvard.edu/abs/2024arXiv240811014S}
}

@ARTICLE{ShishkinSoker2025Crab,
       author = {{Shishkin}, Dmitry and {Soker}, Noam},
        title = "{Et tu, Brute?: The Crab Nebula also exploded by jittering jets}",
      journal = {arXiv e-prints},
         year = 2024,
        month = nov,
          eid = {arXiv:2411.07938},
        pages = {arXiv:2411.07938},
archivePrefix = {arXiv},
       eprint = {2411.07938},
 primaryClass = {astro-ph.HE},
       adsurl = {https://ui.adsabs.harvard.edu/abs/2024arXiv241107938S}
}

@ARTICLE{SokerGilkis2017,
       author = {{Soker}, Noam and {Gilkis}, Avishai},
        title = "{Magnetar-powered Superluminous Supernovae Must First Be Exploded by Jets}",
      journal = {\apj},
         year = 2017,
        month = dec,
       volume = {851},
       number = {2},
          eid = {95},
        pages = {95},
          doi = {10.3847/1538-4357/aa9c83},
archivePrefix = {arXiv},
       eprint = {1708.08356},
 primaryClass = {astro-ph.HE},
       adsurl = {https://ui.adsabs.harvard.edu/abs/2017ApJ...851...95S}
}

@ARTICLE{Kumar2025,
       author = {{Kumar}, Amit},
        title = "{Insights from modelling magnetar-driven light curves of stripped-envelope supernovae}",
      journal = {\na},
         year = 2025,
        month = may,
       volume = {116},
          eid = {102346},
        pages = {102346},
          doi = {10.1016/j.newast.2024.102346},
archivePrefix = {arXiv},
       eprint = {2412.09357},
 primaryClass = {astro-ph.HE},
       adsurl = {https://ui.adsabs.harvard.edu/abs/2025NewA..11602346K}
}

@ARTICLE{Soker2025Learning,
       author = {{Soker}, Noam},
        title = "{Learning from core-collapse supernova remnants on the explosion mechanism}",
      journal = {arXiv e-prints},
         year = 2024,
        month = sep,
          eid = {arXiv:2409.13657},
        pages = {arXiv:2409.13657},
          doi = {10.48550/arXiv.2409.13657},
archivePrefix = {arXiv},
       eprint = {2409.13657},
 primaryClass = {astro-ph.HE},
       adsurl = {https://ui.adsabs.harvard.edu/abs/2024arXiv240913657S}
}

@ARTICLE{Bocciolietal2025,
       author = {{Boccioli}, Luca and {Vartanyan}, David and {O'Connor}, Evan P. and {Kasen}, Daniel},
        title = "{Neutrino heating in 1D, 2D, and 3D core-collapse supernovae: characterizing the explosion of high-compactness stars}",
      journal = {\mnras},
         year = 2025,
        month = jul,
       volume = {540},
       number = {4},
        pages = {3885-3905},
          doi = {10.1093/mnras/staf963},
archivePrefix = {arXiv},
       eprint = {2501.06784},
 primaryClass = {astro-ph.HE},
       adsurl = {https://ui.adsabs.harvard.edu/abs/2025MNRAS.540.3885B}
}

@ARTICLE{Soker2024UnivReview,
       author = {{Soker}, Noam},
        title = "{The Two Alternative Explosion Mechanisms of Core-Collapse Supernovae: 2024 Status Report}",
      journal = {Universe},
         year = 2024,
        month = dec,
       volume = {10},
       number = {12},
          eid = {458},
        pages = {458},
          doi = {10.3390/universe10120458},
archivePrefix = {arXiv},
       eprint = {2411.08555},
 primaryClass = {astro-ph.HE},
       adsurl = {https://ui.adsabs.harvard.edu/abs/2024Univ...10..458S}
}

@ARTICLE{Bearetal2025Puppis,
       author = {{Bear}, Ealeal and {Shishkin}, Dmitry and {Soker}, Noam},
        title = "{The Puppis A Supernova Remnant: An Early Jet-driven Neutron Star Kick followed by Jittering Jets}",
      journal = {Research in Astronomy and Astrophysics},
         year = 2025,
        month = apr,
       volume = {25},
       number = {4},
          eid = {045008},
        pages = {045008},
          doi = {10.1088/1674-4527/adc24e},
archivePrefix = {arXiv},
       eprint = {2409.11453},
 primaryClass = {astro-ph.HE},
       adsurl = {https://ui.adsabs.harvard.edu/abs/2025RAA....25d5008B}
}

@ARTICLE{Moriyaetal2025,
       author = {{Moriya}, Takashi J. and {Coulter}, David A. and {DeCoursey}, Christa and {Pierel}, Justin D.~R. and {Hainline}, Kevin and {Siebert}, Matthew R. and {Rest}, Armin and {Egami}, Eiichi and {Gomez}, Sebastian and {Quimby}, Robert M. and {Fox}, Ori D. and {Engesser}, Michael and {Sun}, Fengwu and {Chen}, Wenlei and {Zenati}, Yossef and {Gezari}, Suvi and {Joshi}, Bhavin A. and {Shahbandeh}, Melissa and {Strolger}, Louis-Gregory and {Wang}, Qinan and {Alberts}, Stacey and {Bhatawdekar}, Rachana and {Bunker}, Andrew J. and {Rinaldi}, Pierluigi and {Robertson}, Brant E. and {Tacchella}, Sandro},
        title = "{Properties of high-redshift Type II supernovae discovered by the JADES transient survey}",
      journal = {arXiv e-prints},
         year = 2025,
        month = jan,
          eid = {arXiv:2501.08969},
        pages = {arXiv:2501.08969},
archivePrefix = {arXiv},
       eprint = {2501.08969},
 primaryClass = {astro-ph.HE},
       adsurl = {https://ui.adsabs.harvard.edu/abs/2025arXiv250108969M}
}

@ARTICLE{BoccioliFragione2024,
       author = {{Boccioli}, Luca and {Fragione}, Giacomo},
        title = "{Remnant masses from 1 D + core-collapse supernovae simulations: Bimodal neutron star mass distribution and black holes in the low-mass gap}",
      journal = {\prd},
         year = 2024,
        month = jul,
       volume = {110},
       number = {2},
          eid = {023007},
        pages = {023007},
          doi = {10.1103/PhysRevD.110.023007},
archivePrefix = {arXiv},
       eprint = {2404.05927},
 primaryClass = {astro-ph.HE},
       adsurl = {https://ui.adsabs.harvard.edu/abs/2024PhRvD.110b3007B}
}

@ARTICLE{StrotjohannOfekGalYam2024,
       author = {{Strotjohann}, Nora L. and {Ofek}, Eran O. and {Gal-Yam}, Avishay},
        title = "{A Bias-corrected Luminosity Function for Red Supergiant Supernova Progenitor Stars}",
      journal = {\apjl},
         year = 2024,
        month = apr,
       volume = {964},
       number = {2},
          eid = {L27},
        pages = {L27},
          doi = {10.3847/2041-8213/ad3064},
archivePrefix = {arXiv},
       eprint = {2311.00744},
 primaryClass = {astro-ph.HE},
       adsurl = {https://ui.adsabs.harvard.edu/abs/2024ApJ...964L..27S}
}

@ARTICLE{ByrneFraser2022,
       author = {{Byrne}, R.~A. and {Fraser}, M.},
        title = "{Nothing to see here: failed supernovae are faint or rare}",
      journal = {\mnras},
         year = 2022,
        month = jul,
       volume = {514},
       number = {1},
        pages = {1188-1205},
          doi = {10.1093/mnras/stac1308},
archivePrefix = {arXiv},
       eprint = {2201.12187},
 primaryClass = {astro-ph.SR},
       adsurl = {https://ui.adsabs.harvard.edu/abs/2022MNRAS.514.1188B}
}

@ARTICLE{Healyetal2025,
       author = {{Healy}, Sarah and {Horiuchi}, Shunsaku and {Ashall}, Chris},
        title = "{The Red Supergiant Problem: As Seen from the Local Group's Red Supergiant Populations}",
      journal = {arXiv e-prints},
         year = 2024,
        month = dec,
          eid = {arXiv:2412.04386},
        pages = {arXiv:2412.04386},
archivePrefix = {arXiv},
       eprint = {2412.04386},
 primaryClass = {astro-ph.SR},
       adsurl = {https://ui.adsabs.harvard.edu/abs/2024arXiv241204386H}
}

@ARTICLE{BeasoretalLuminosity2025,
       author = {{Beasor}, Emma R. and {Smith}, Nathan and {Jencson}, Jacob E.},
        title = "{The Red Supergiant Progenitor Luminosity Problem}",
      journal = {\apj},
         year = 2025,
        month = feb,
       volume = {979},
       number = {2},
          eid = {117},
        pages = {117},
          doi = {10.3847/1538-4357/ad8f3f},
archivePrefix = {arXiv},
       eprint = {2410.14027},
 primaryClass = {astro-ph.SR},
       adsurl = {https://ui.adsabs.harvard.edu/abs/2025ApJ...979..117B}
}

@ARTICLE{Shibataetal2025,
       author = {{Shibata}, Masaru and {Fujibayashi}, Sho and {Wanajo}, Shinya and {Ioka}, Kunihito and {Lam}, Alan Tsz-Lok and {Sekiguchi}, Yuichiro},
        title = "{Self-consistent scenario for jet and stellar explosions in collapsar: General relativistic magnetohydrodynamics simulation with a dynamo}",
      journal = {\prd},
         year = 2025,
        month = jun,
       volume = {111},
       number = {12},
          eid = {123017},
        pages = {123017},
          doi = {10.1103/msy2-fwhx},
archivePrefix = {arXiv},
       eprint = {2502.02077},
 primaryClass = {astro-ph.HE},
       adsurl = {https://ui.adsabs.harvard.edu/abs/2025PhRvD.111l3017S}
}

@ARTICLE{Imashevaetal2025,
       author = {{Imasheva}, Liliya and {Janka}, H. -Thomas and {Weiss}, Achim},
        title = "{Comparison of Three Methods for Triggering Core-collapse Supernova Explosions in Spherical Symmetry}",
      journal = {arXiv e-prints},
         year = 2025,
        month = jan,
          eid = {arXiv:2501.13172},
        pages = {arXiv:2501.13172},
          doi = {10.48550/arXiv.2501.13172},
archivePrefix = {arXiv},
       eprint = {2501.13172},
 primaryClass = {astro-ph.HE},
       adsurl = {https://ui.adsabs.harvard.edu/abs/2025arXiv250113172I}
}

@ARTICLE{SokerShishkin2025Vela,
       author = {{Soker}, Noam and {Shishkin}, Dmitry},
        title = "{The Vela Supernova Remnant: The Unique Morphological Features of Jittering Jets}",
      journal = {Research in Astronomy and Astrophysics},
         year = 2025,
        month = mar,
       volume = {25},
       number = {3},
          eid = {035008},
        pages = {035008},
          doi = {10.1088/1674-4527/adb4cc},
archivePrefix = {arXiv},
       eprint = {2409.02626},
 primaryClass = {astro-ph.HE},
       adsurl = {https://ui.adsabs.harvard.edu/abs/2025RAA....25c5008S}
}

@ARTICLE{ShishkinSoker2025G321,
       author = {{Shishkin}, Dmitry and {Soker}, Noam},
        title = "{The point-symmetrical morphology of supernova remnant G321.3-3.9: Exploding jets jitter around an axis}",
      journal = {In preparation},
         year = 2025,
        month = Jul,
       volume = {},
       number = {},
        pages = {},
          doi = {},
archivePrefix = {},
       eprint = {},
 primaryClass = {astro-ph.HE},
       adsurl = {}
}

@ARTICLE{Soker2024W44,
       author = {{Soker}, Noam},
        title = "{Identifying a Point-Symmetrical Morphology in the Core-Collapse Supernova Remnant W44}",
      journal = {Universe},
         year = 2024,
        month = dec,
       volume = {11},
       number = {1},
          eid = {4},
        pages = {4},
          doi = {10.3390/universe11010004},
archivePrefix = {arXiv},
       eprint = {2411.04654},
 primaryClass = {astro-ph.HE},
       adsurl = {https://ui.adsabs.harvard.edu/abs/2024Univ...11....4S}
}

@ARTICLE{BearSoker2025,
       author = {{Bear}, Ealeal and {Soker}, Noam},
        title = "{Identifying a point-symmetric morphology in supernova remnant Cassiopeia A: Explosion by jittering jets}",
      journal = {\na},
         year = 2025,
        month = jan,
       volume = {114},
          eid = {102307},
        pages = {102307},
          doi = {10.1016/j.newast.2024.102307},
archivePrefix = {arXiv},
       eprint = {2403.07625},
 primaryClass = {astro-ph.HE},
       adsurl = {https://ui.adsabs.harvard.edu/abs/2025NewA..11402307B}
}

@ARTICLE{Soker2025G0901,
       author = {{Soker}, Noam},
        title = "{The morphology of supernova remnant G0.9+0.1 implies explosion by jittering-jets}",
      journal = {arXiv e-prints},
         year = 2025,
        month = apr,
          eid = {arXiv:2504.11384},
        pages = {arXiv:2504.11384},
          doi = {10.48550/arXiv.2504.11384},
archivePrefix = {arXiv},
       eprint = {2504.11384},
 primaryClass = {astro-ph.HE},
       adsurl = {https://ui.adsabs.harvard.edu/abs/2025arXiv250411384S}
}

@ARTICLE{Laplaceetal2025,
       author = {{Laplace}, E. and {Schneider}, F.~R.~N. and {Podsiadlowski}, Ph.},
        title = "{It's written in the massive stars: The role of stellar physics in the formation of black holes}",
      journal = {\aap},
         year = 2025,
        month = mar,
       volume = {695},
          eid = {A71},
        pages = {A71},
          doi = {10.1051/0004-6361/202451077},
archivePrefix = {arXiv},
       eprint = {2409.02058},
 primaryClass = {astro-ph.SR},
       adsurl = {https://ui.adsabs.harvard.edu/abs/2025A&A...695A..71L}
}

@ARTICLE{Huangetal2025,
       author = {{Huang}, Xu-Run and {Zha}, Shuai and {Chu}, Ming-chung and {O'Connor}, Evan P. and {Chen}, Lie-Wen},
        title = "{Phase-transition-induced Collapse of Proto-compact Stars and Its Implication for Supernova Explosions}",
      journal = {\apj},
         year = 2025,
        month = feb,
       volume = {979},
       number = {2},
          eid = {151},
        pages = {151},
          doi = {10.3847/1538-4357/ada146},
archivePrefix = {arXiv},
       eprint = {2409.16189},
 primaryClass = {astro-ph.HE},
       adsurl = {https://ui.adsabs.harvard.edu/abs/2025ApJ...979..151H}
}

@ARTICLE{EggenbergerAndersenetal2025,
       author = {{Eggenberger Andersen}, Oliver and {O'Connor}, Evan and {Andresen}, Haakon and {da Silva Schneider}, Andr{\'e} and {Couch}, Sean M.},
        title = "{Black Hole Supernovae, Their Equation of State Dependence, and Ejecta Composition}",
      journal = {\apj},
         year = 2025,
        month = feb,
       volume = {980},
       number = {1},
          eid = {53},
        pages = {53},
          doi = {10.3847/1538-4357/ada899},
archivePrefix = {arXiv},
       eprint = {2411.11969},
 primaryClass = {astro-ph.HE},
       adsurl = {https://ui.adsabs.harvard.edu/abs/2025ApJ...980...53E}
}

@ARTICLE{Maltsevetal2025,
       author = {{Maltsev}, K. and {Schneider}, F.~R.~N. and {Mandel}, I. and {M{\"u}ller}, B. and {Heger}, A. and {R{\"o}pke}, F.~K. and {Laplace}, E.},
        title = "{Explodability criteria for the neutrino-driven supernova mechanism}",
      journal = {arXiv e-prints},
         year = 2025,
        month = mar,
          eid = {arXiv:2503.23856},
        pages = {arXiv:2503.23856},
          doi = {10.48550/arXiv.2503.23856},
archivePrefix = {arXiv},
       eprint = {2503.23856},
 primaryClass = {astro-ph.SR},
       adsurl = {https://ui.adsabs.harvard.edu/abs/2025arXiv250323856M}
}

@ARTICLE{Maunderetal2025,
       author = {{Maunder}, Thomas and {Callan}, Fionntan P. and {Sim}, Stuart A. and {Heger}, Alexander and {M{\"u}ller}, Bernhard},
        title = "{Synthetic Light Curves and Spectra for the Photospheric Phase of a 3D Stripped-Envelope Supernova Explosion Model}",
      journal = {arXiv e-prints},
         year = 2024,
        month = oct,
          eid = {arXiv:2410.20829},
        pages = {arXiv:2410.20829},
          doi = {10.48550/arXiv.2410.20829},
archivePrefix = {arXiv},
       eprint = {2410.20829},
 primaryClass = {astro-ph.HE},
       adsurl = {https://ui.adsabs.harvard.edu/abs/2024arXiv241020829M}
}

@ARTICLE{Mulleretal2025,
       author = {{M{\"u}ller}, Bernhard and {Heger}, Alexander and {Powell}, Jade},
        title = "{Minimum Neutron Star Mass in Neutrino-Driven Supernova Explosions}",
      journal = {\prl},
         year = 2025,
        month = feb,
       volume = {134},
       number = {7},
          eid = {071403},
        pages = {071403},
          doi = {10.1103/PhysRevLett.134.071403},
archivePrefix = {arXiv},
       eprint = {2407.08407},
 primaryClass = {astro-ph.HE},
       adsurl = {https://ui.adsabs.harvard.edu/abs/2025PhRvL.134g1403M}
}

@ARTICLE{WangBurrows2025,
       author = {{Wang}, Tianshu and {Burrows}, Adam},
        title = "{The Effect of the Fast-Flavor Instability on Core-Collapse Supernova Models}",
      journal = {arXiv e-prints},
         year = 2025,
        month = mar,
          eid = {arXiv:2503.04896},
        pages = {arXiv:2503.04896},
          doi = {10.48550/arXiv.2503.04896},
archivePrefix = {arXiv},
       eprint = {2503.04896},
 primaryClass = {astro-ph.HE},
       adsurl = {https://ui.adsabs.harvard.edu/abs/2025arXiv250304896W}
}

@ARTICLE{SykesMuller2025,
       author = {{Sykes}, Bailey and {M{\"u}ller}, Bernhard},
        title = "{Long-time 3D supernova simulations of nonrotating progenitors with magnetic fields}",
      journal = {\prd},
         year = 2025,
        month = mar,
       volume = {111},
       number = {6},
          eid = {063042},
        pages = {063042},
          doi = {10.1103/PhysRevD.111.063042},
archivePrefix = {arXiv},
       eprint = {2412.01155},
 primaryClass = {astro-ph.HE},
       adsurl = {https://ui.adsabs.harvard.edu/abs/2025PhRvD.111f3042S}
}

@ARTICLE{Janka2025,
       author = {{Janka}, H. -Thomas},
        title = "{Long-Term Multidimensional Models of Core-Collapse Supernovae: Progress and Challenges}",
      journal = {arXiv e-prints},
         year = 2025,
        month = feb,
          eid = {arXiv:2502.14836},
        pages = {arXiv:2502.14836},
          doi = {10.48550/arXiv.2502.14836},
archivePrefix = {arXiv},
       eprint = {2502.14836},
 primaryClass = {astro-ph.HE},
       adsurl = {https://ui.adsabs.harvard.edu/abs/2025arXiv250214836J}
}

@ARTICLE{ParadisoCoughlin2025,
       author = {{Paradiso}, Daniel A. and {Coughlin}, Eric R.},
        title = "{Gotta Go Fast: A Generalization of the Escape Speed to Fluid-dynamical Explosions and Implications for Astrophysical Transients}",
      journal = {arXiv e-prints},
         year = 2025,
        month = apr,
          eid = {arXiv:2504.11527},
        pages = {arXiv:2504.11527},
archivePrefix = {arXiv},
       eprint = {2504.11527},
 primaryClass = {astro-ph.HE},
       adsurl = {https://ui.adsabs.harvard.edu/abs/2025arXiv250411527P}
}

@ARTICLE{Hortonetal2025,
       author = {{Horton}, M.~A. and {Hardcastle}, M.~J. and {Miley}, G.~K. and {Tasse}, C. and {Shimwell}, T.},
        title = "{Complex morphology and precession indicators of active galactic nuclei jets in LoTSS DR2}",
      journal = {\aap},
         year = 2025,
        month = jul,
       volume = {699},
          eid = {A338},
        pages = {A338},
          doi = {10.1051/0004-6361/202453559},
archivePrefix = {arXiv},
       eprint = {2504.18518},
 primaryClass = {astro-ph.GA},
       adsurl = {https://ui.adsabs.harvard.edu/abs/2025A&A...699A.338H}
}

@ARTICLE{Chuetal1987,
       author = {{Chu}, You-Hua and {Jacoby}, George H. and {Arendt}, Richard},
        title = "{Multiple-Shell Planetary Nebulae. I. Morphologies and Frequency of Occurrence}",
      journal = {\apjs},
         year = 1987,
        month = jul,
       volume = {64},
        pages = {529},
          doi = {10.1086/191207},
       adsurl = {https://ui.adsabs.harvard.edu/abs/1987ApJS...64..529C}
}

@ARTICLE{Sahaietal2007,
       author = {{Sahai}, Raghvendra and {Morris}, Mark and {S{\'a}nchez Contreras}, Carmen and {Claussen}, Mark},
        title = "{Preplanetary Nebulae: A Hubble Space Telescope Imaging Survey and a New Morphological Classification System}",
      journal = {\aj},
         year = 2007,
        month = dec,
       volume = {134},
       number = {6},
        pages = {2200-2225},
          doi = {10.1086/522944},
archivePrefix = {arXiv},
       eprint = {0707.4662},
 primaryClass = {astro-ph},
       adsurl = {https://ui.adsabs.harvard.edu/abs/2007AJ....134.2200S}
}

@ARTICLE{Sahaietal2011,
       author = {{Sahai}, Raghvendra and {Morris}, Mark R. and {Villar}, Gregory G.},
        title = "{Young Planetary Nebulae: Hubble Space Telescope Imaging and a New Morphological Classification System}",
      journal = {\aj},
         year = 2011,
        month = apr,
       volume = {141},
       number = {4},
          eid = {134},
        pages = {134},
          doi = {10.1088/0004-6256/141/4/134},
archivePrefix = {arXiv},
       eprint = {1101.2214},
 primaryClass = {astro-ph.GA},
       adsurl = {https://ui.adsabs.harvard.edu/abs/2011AJ....141..134S}
}

@ARTICLE{Balick1987,
       author = {{Balick}, Bruce},
        title = "{The Evolution of Planetary Nebulae. I. Structures, Ionizations, and Morphological Sequences}",
      journal = {\aj},
         year = 1987,
        month = sep,
       volume = {94},
        pages = {671},
          doi = {10.1086/114504},
       adsurl = {https://ui.adsabs.harvard.edu/abs/1987AJ.....94..671B}
}

@ARTICLE{Braudoetal2025,
       author = {{Braudo}, Jessica and {Michaelis}, Amir and {Akashi}, Muhammad and {Soker}, Noam},
        title = "{Simulating the Shaping of Point-symmetric Structures in the Jittering Jets Explosion Mechanism}",
      journal = {\pasp},
         year = 2025,
        month = may,
       volume = {137},
       number = {5},
          eid = {054201},
        pages = {054201},
          doi = {10.1088/1538-3873/add08e},
archivePrefix = {arXiv},
       eprint = {2503.10326},
 primaryClass = {astro-ph.HE},
       adsurl = {https://ui.adsabs.harvard.edu/abs/2025PASP..137e4201B}
}

@ARTICLE{Izzoetal2019,
       author = {{Izzo}, L. and {de Ugarte Postigo}, A. and {Maeda}, K. and {Th{\"o}ne}, C.~C. and {Kann}, D.~A. and {Della Valle}, M. and {Sagues Carracedo}, A. and {Micha{\l}owski}, M.~J. and {Schady}, P. and {Schmidl}, S. and {Selsing}, J. and {Starling}, R.~L.~C. and {Suzuki}, A. and {Bensch}, K. and {Bolmer}, J. and {Campana}, S. and {Cano}, Z. and {Covino}, S. and {Fynbo}, J.~P.~U. and {Hartmann}, D.~H. and {Heintz}, K.~E. and {Hjorth}, J. and {Japelj}, J. and {Kami{\'n}ski}, K. and {Kaper}, L. and {Kouveliotou}, C. and {Kru{\.Z}y{\'n}ski}, M. and {Kwiatkowski}, T. and {Leloudas}, G. and {Levan}, A.~J. and {Malesani}, D.~B. and {Micha{\l}owski}, T. and {Piranomonte}, S. and {Pugliese}, G. and {Rossi}, A. and {S{\'a}nchez-Ram{\'\i}rez}, R. and {Schulze}, S. and {Steeghs}, D. and {Tanvir}, N.~R. and {Ulaczyk}, K. and {Vergani}, S.~D. and {Wiersema}, K.},
        title = "{Signatures of a jet cocoon in early spectra of a supernova associated with a {\ensuremath{\gamma}}-ray burst}",
      journal = {\nat},
         year = 2019,
        month = jan,
       volume = {565},
       number = {7739},
        pages = {324-327},
          doi = {10.1038/s41586-018-0826-3},
archivePrefix = {arXiv},
       eprint = {1901.05500},
 primaryClass = {astro-ph.HE},
       adsurl = {https://ui.adsabs.harvard.edu/abs/2019Natur.565..324I}
}

@ARTICLE{Rhodes2006,
       author = {{Rhodes}, Gillianr},
        title = "{THE EVOLUTIONARY PSYCHOLOGY OF FACIAL BEAUTY}",
      journal = {Annu. Rev. Psychol},
         year = 2006,
        month = Jan,
       volume = {57},
       number = {2},
          eid = {199},
        pages = {199},
          doi = {10.1146/annurev.psych.57.102904.190208},
archivePrefix = {arXiv},
       eprint = {0000.000000},
 primaryClass = {astro-ph.HE},
       adsurl = {}
}

@ARTICLE{Pinheiroetal2023,
       author = {{Pinheiro}, Luiza Penha and {Monteiro}, Luis Carlos Pereira and {Henriques}, Leonardo Dutra and {Souza}, Givago Silva and {Miranda}, Ana Catarina and {Costa}, Marcelo Fernandes and {Henriques}, Alda Loureiro},
        title = "{Association between Facial Metrics and Mate Rejection for Long-Term Relationship by Heterosexual Men}",
      journal = {Symmetry},
         year = 2023,
        month = jan,
       volume = {15},
       number = {1},
          eid = {133},
        pages = {133},
          doi = {10.3390/sym15010133},
       adsurl = {https://ui.adsabs.harvard.edu/abs/2023Symm...15..133P}
}
  \bibliographystyle{aasjournal}

\end{document}